\documentclass{aa}

\usepackage{txfonts}
\usepackage{graphicx}

\usepackage{amsmath}    
\usepackage{amssymb}    

\usepackage{soul}
\usepackage{ulem}

\usepackage{hyperref}

\begin{document}

\title{Impact of non-gravitational effects on chaotic properties of retrograde orbits}

\authorrunning{Kankiewicz \& W\l odarczyk}

\titlerunning{Non-gravitational effects in retrograde orbits}

\author{Pawe{\l} Kankiewicz$^{1}$, Ireneusz W\l odarczyk$^{2}$}

 \institute{Institute of Physics, Astrophysics Division, Jan Kochanowski University, Swietokrzyska 15, 25-406 Kielce, Poland
 \and
Chorz\'ow Astronomical Observatory (MPC 553), Al. Planetarium 4, 41-500 Chorz\'ow, Poland
}
         
\date{\today} 

\offprints{P. Kankiewicz, \\ e-mail: {\tt pawel.kankiewicz@ujk.edu.pl}}

   \date{Received xx xx xxxx / Accepted xx xx xxxx} 
   
 \abstract
 {Dynamical studies of asteroid populations in retrograde orbits, that is with orbital inclinations greater than 90 degrees, are interesting because the origin of such orbits is still unexplained. Generally, the population of retrograde asteroids includes mostly Centaurs and transneptunian objects (TNOs). A special case is the near-Earth object (343158) 2009 HC82 from the Apollo group. Another interesting object is the comet 333P/LINEAR, which for several years was considered the second retrograde object approaching Earth. Another comet in retrograde orbit, 161P Hartley/IRAS appears to be an object of similar type. Thanks to the large amount of observational data for these two comets, we tested various models of cometary non-gravitational forces applied to their dynamics.}
 {The goal was to estimate which of non-gravitational perturbations could affect the stability of retrograde bodies. In principle, we study the local stability by measuring the divergence of nearby orbits.}
 {We numerically determined Lyapunov characteristic indicators (LCI) and the associated Lyapunov times (LT). This time, our calculations of these parameters were extended by more advanced models of non-gravitational perturbations (i.e. Yarkovsky drift and in selected cases cometary forces). This allowed us to estimate chaos in the Lyapunov sense. 
 }
{We found that the Yarkovsky effect for obliquities of $\gamma=0^{\circ}$ and $\gamma=180^{\circ}$ can change the LT substantially. In most cases, for the prograde rotation, we received more stable solutions. Moreover, we confirmed the role of retrograde resonances in this process. Additionally, the studied cometary effects also significantly influence the long-term behaviour of the selected comets. The LT can reach values from 100 to over 1000 years. 
}
{All of our results indicate that the use of models with non-gravitational effects for retrograde bodies is clearly justified.
}


\maketitle

\section{Introduction}
Currently, a small number of retrograde asteroids are known, mainly among transneptunian objects (TNOs) and Centaurs. Their puzzling existence is an intriguing problem, which has yet to be thoroughly explained.
In principle, we define a retrograde asteroid as a small body with an orbital inclination greater than 90 degrees. It is a common opinion that such orbits are typical of comets, so the confirmation of cometary activity for a fraction of these objects is possible in the near future. 
About 18\% of the population of bright comets have orbits with large inclinations (between 90 and 180 degrees). Comparatively, among the known orbits of about 995,000 asteroids (September 2020), only 117 are now classified as retrograde. 
We can assume that the retrograde orbital motion in the Solar System is typical for comets, but the existence of a small population of retrograde asteroids is still unexplained.
The key question is whether these asteroids are just inactive comets or if they are a completely different class of small bodies.
Many discovered retrograde bodies still retain their formal classification as asteroids. Moreover, more exotic objects have been discovered in recent years, such as hyperbolic asteroids and a retrograde Trojan, trapped in a 1:1 resonance with Jupiter \citep{Wiegert2017}.

The main aim of our studies was to estimate the role of non-gravitational forces in the dynamics of retrograde asteroids.
In particular, we were interested in possible chaotic behaviour due to these forces, which was estimated quantitatively by chaotic indicators. Because the presence of chaos possibly dominates the migration of these bodies, we checked the influence of these small forces on chaotic properties.

This work is a continuation of our long-term project on the dynamics of retrograde objects. In previous studies, we determined various quantitative parameters related to the long-term orbital stability of retrograde asteroids. Initially, these were the median dynamical lifetimes ($\tau$), which determined the rate of removal of retrograde test particles from the Solar System. The term ``removal'' is understood in this work primarily as ejections into distant orbits, but also as collisions with the Sun or planets. As the example of the retrograde asteroid 514107 (2015 BZ509), described by \cite{Namouni2020}, shows the cases of collisions with the Sun of test particles may be more frequent than the ejections.

Our first estimation of $\tau$ values for retrograde orbits is described in detail in \cite{Kankiewicz2017}. In this study, we found that this parameter is sensitive to relatively weak non-gravitational forces, such as the Yar\-kov\-sky effect. The simplified model showed that this effect could slightly lengthen the $\tau$ values. Another interesting conclusion was the comparison of backward and forward integration. Rapid changes in the orbital elements were more frequent in the past than in the future. The resulting $\tau$ values were also different in the past and the future. A possible explanation is the existence of different dynamical regimes (chaotic zones) in orbital evolution. To verify the previously obtained parameters and confirm the possible presence of chaos, we supplemented these results with the Lyapunov exponents of these orbits, which we estimated with a simplified gravitational model \citep{Kankiewicz2018}. 
The general conclusion of this study is that $\tau$ and Lyapunov times (LT) allow for a quantitative estimation of dependence on
initial conditions, but are not necessarily strongly correlated. For example, the exceptions were objects with the most extended lifetimes. Similar cases are often described as ``stable chaos'' \citep{Milani1992}. Lyapunov times are based on the separation of close trajectories in six-dimensional parameter space, so we chose these for further use as more universal in application. In contrast to the previous work, where the term ``stability'' usually referred to dynamical stability (the body remaining inside the system), we used stability in the sense of Lyapunov. It means a quantitative estimation of sensitivity to the initial conditions, that is chaos in a mathematical sense. Therefore, the priority is to study the predictability of orbits, depending on the model of possible non-gravitational perturbations used.

The last, meaningful comment in the summary of our previous work was the unresolved issue of cometary non-gravitational effects. In this current study, we decided to develop these considerations by extending our research to more complex models of possible non-gravitational forces. In addition to more advanced models, we also made a significant update of the observational data.

\begin{figure}
\includegraphics[width=1.0\columnwidth]{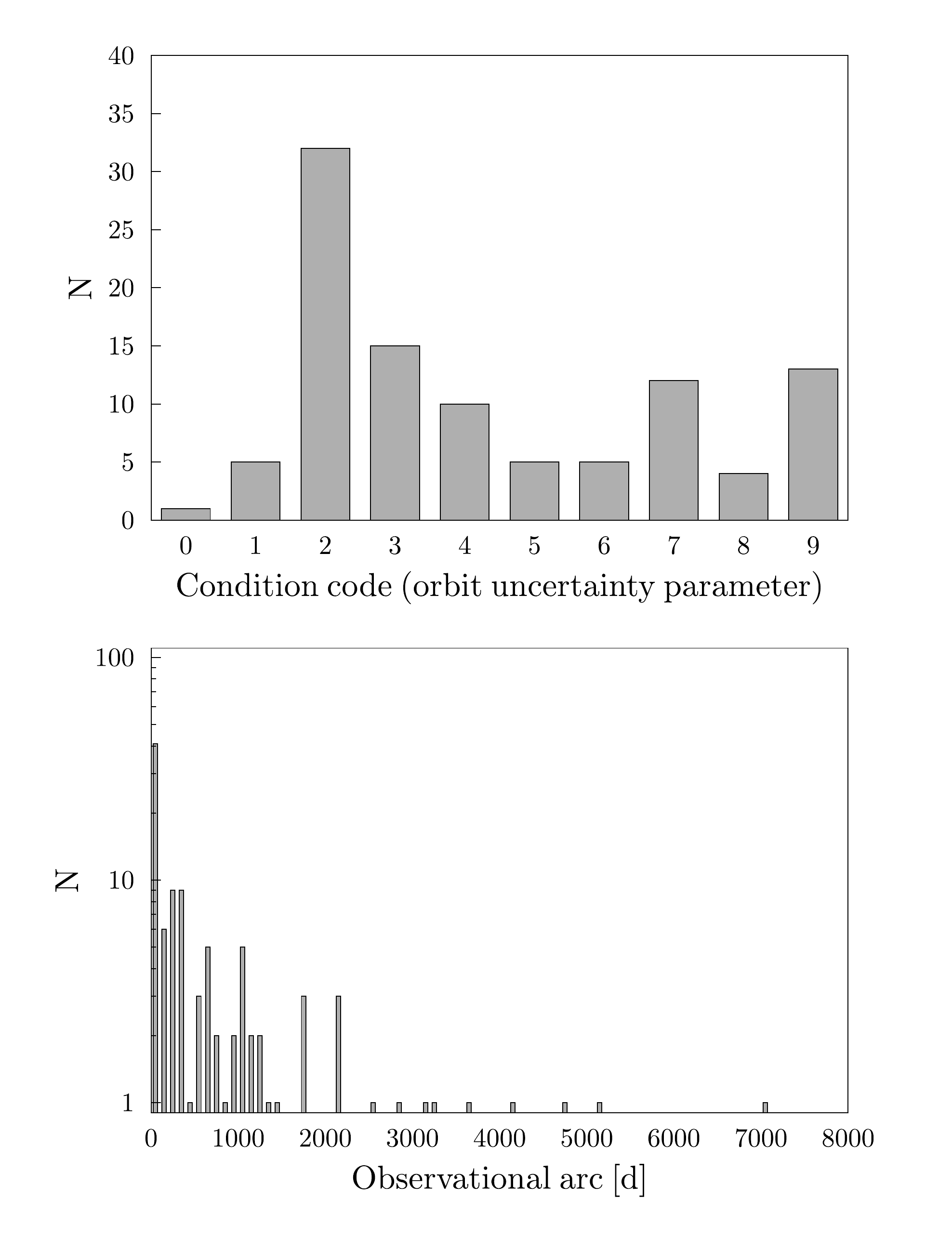}
\caption{Observational statistics of retrograde asteroids: MPC orbit uncertainties (``U'' parameters) and the length of observational arcs in days.}
\label{fig1}
\end{figure}

Although there are 105 catalogued retrograde asteroids available, many orbits are determined with low precision. Observational arcs are short and some objects were observed only for a few days. This fact caused problems with the propagation of observational errors, which limited reliable estimates of the lifetime and stability of orbits. Consequently, we opted for multi-oppositional asteroids only. So far, we consistently worked on the same sample of 25 ``old'' and 6 ``new'' asteroids, making only minor updates to the orbital elements when the orbits were corrected. According to the latest observational statistics, about 35\% of objects have acceptable uncertainty codes and sufficiently long observational arcs (Fig. \ref{fig1}). 

There are several different scenarios for the origin of the retrograde orbit. First of all, the number of these orbits is small, which leads to the preliminary conclusion that their lifetimes must be short. All the results indicate that these values are of the order of millions of years, and their production mechanism must hardly be effective. Another potential explanation may be the prograde rotation of the protoplanetary disc, which causes the accidental inversions of orbits to be extremely rare.

Because there are a lot of different concepts for the creation of orbits with large inclinations, various alternative scenarios for the formation of retrograde asteroids cannot be excluded. An interesting fact is the significant presence of retrograde meteorites. Because collisions are not uncommon in this type of dynamical processes, these objects can be produced by impacts. \cite{Greenstreet2012} described this in more detail in their scenario: 
retrograde near-Earth asteroids (NEAs), which are produced by resonances and migrated from the main melt, are subject to collisions, creating retrograde debris. 
It is possible that the retrograde meteorites originate from the main belt. A similar scenario concerning meteorites has been proposed by \cite{Wiegert2017a}. According to this concept, collisions of small bodies (including moons of planets) can generate a transfer to unusual orbits that can feed a population of retrograde meteorites. Thus, it is possible to expect that they contain matter imported from other bodies of the Solar System.

An alternative idea is the interstellar origin of retrograde asteroids, which was suggested by \cite{Namouni2018}. More precisely, it is the capture of a small body from outside into the planetary system. The particular example shown in this work showed that the probability of such an event is significant. Moreover, the latest simulations of \cite{Marceta2020} indicate that there are a very large number of retrograde orbits among interstellar objects. This is a serious clue to further exploration in this area.

So far, there are a number of dynamical events in the Solar System that affect the evolution of retrograde orbits. Several orbital resonances may play an important role in this process. Some Centaurs and Damocloids have been captured in retrograde resonances with Jupiter and Saturn, as described in detail by \cite{Morais2013}. The efficiency of retrograde resonances at orbital capture was subsequently explained by \cite{Namouni2017}.

There are also mean motion resonances (MMRs), such as 3:1 with Jupiter, which can possibly produce NEA orbits with high inclinations, including retrograde \citep{Greenstreet2012}. A similar mechanism, connected with resonances of a higher order, was suggested by \cite{Marcos2014}. Additionally, the scenario of orbit ``flipping'' resulting from the influence of the Kozai resonance, was proposed by \cite{Marcos2015} and explained in the particular example of comet 96P/Machholz 1. Recently, all significant retrograde MMRs have been classified in the \cite{Li2019} studies. In that work, the authors analyse 143 resonance configurations that may have retrograde asteroids. Many new resonances were identified, corresponding to objects from our list.

\section{Data and methods}
\subsection{Initial data}

To avoid cases with poorly determined orbits in our calculations, the sample of 31 multi-opposition retrograde asteroids was used (Tab. \ref{object_list}). In our recent study \citep{Kankiewicz2017}, we analysed the dynamical lifetimes of the same asteroid group (not including 6 asteroids added later to this paper after the update of observational data). All of the objects from this list were classified as either Centaurs, TNOs, or NEAs (only one Apollo-type asteroid). Currently, there are only 14 numbered asteroids in this population. Many of other retrograde objects are still ``border cases'' of minor bodies that have potential cometary activity. 
\begin{table}
\footnotesize\addtolength{\tabcolsep}{-3pt}
\caption{List of 31 numbered and multi-opposition retrograde asteroids with $i$, $a$, $e$, observational arc in days and the number of observations used in the orbit determination. All data adopted from JPL Small-Body Database. Information updated since our previous studies \citep{Kankiewicz2018} is denoted with an asterisk. New objects in the list are denoted with a double asterisk.\label{object_list}}
\begin{tabular}{|c|c|c|c|c|c|c|}
\hline
N & Name & $i$ [deg] & $a$ [au] & $e$ & arc [d] & $N_{obs}$ \\
\hline
1& (20461) Dioretsa & 160.4&23.9&0.900&927&277\\ 
2& (65407) 2002 RP120 & 119.0&53.9&0.954&1226&536\\ 
3& {(330759) 2008 SO218}$^{\star}$& 170.3&8.1&0.564&3132&348\\ 
4& {(336756) 2010 NV1}$^{\star}$ & 140.7&290.1&0.968&2143&157\\
5& (342842) 2008 YB3 & 105.1&11.6&0.441&3236&502\\ 
6& (343158) 2009 HC82 & 154.4&2.5&0.807&4794&163\\ 
7& (434620) 2005 VD & 172.9&6.7&0.252&6222&60\\ 
8& (459870) 2014 AT28 & 165.6&10.9&0.404&2362&79\\ 

9& {1999 LE31}$^{\star}$ & 151.8&8.1&0.466&7022&130\\ 
10& 2000 HE46 & 158.5&23.6&0.899&220&119\\ 
11& 2004 NN8 & 165.5&98.4&0.976&292&82\\ 
12& 2005 SB223 & 91.3&29.4&0.905&244&50\\ 
13& 2006 BZ8 & 165.3&9.6&0.803&623&345\\ 
14& 2008 KV42 & 103.4&41.7&0.493&3610&58\\ 
15& 2009 QY6 & 137.7&12.5&0.834&264&167\\ 
16& 2009 YS6 & 147.8&20.1&0.920&1271&157\\ 
17& 2010 CG55 & 146.2&31.9&0.910&1075&124\\ 
18& 2010 GW64 & 105.2&63.1&0.941&252&38\\ 
19& 2010 OR1 & 143.9&26.9&0.924&232&152\\ 

20& 2011 MM4 &100.5&21.13&0.473&1071&111 \\
21& 2012 HD2 &146.9&62.39&0.959&302&107 \\
22& 2013 BL76 &98.61&1036&0.992&687&70 \\
23& 2013 LD16 &154.8&79.99&0.968&262&54 \\
24& (468861) 2013 LU28 &125.4&175.5&0.950&1805&110 \\
25& 2013 NS11 &130.3&12.65&0.787&565&257 \\
26 &{(471325) 2011 KT19$^{\star \star}$} & 110.1&35.6&0.330&1779&132\\
27 &{(514107) 2015 BZ509$^{\star \star}$} & 163.0&5.1&0.380&1026&73\\ 
28 &{(518151) 2016 FH13$^{\star \star}$} & 93.6&24.5&0.614&1478&87\\ 
29 &{(523797) 2016 NM56$^{\star \star}$} & 144.0&72.7&0.855&2142&107\\
30 &{(523800) 2017 KZ31$^{\star \star}$} & 161.7&53.4&0.795&1715&88\\ 
31 & {2005 NP82$^{\star \star}$} & 130.5&5.9&0.478&5137&132\\ 
\hline
\end{tabular}
\end{table}

\subsection{Lyapunov times estimation}

A practical and popular tool for the quantitative estimation of chaos are indicators based on Lyapunov exponents. These indicators are often used in the dynamics of asteroids. Parameters such as Lyapunov exponents allow for the quantitative estimation of the divergence of close trajectories.
Lyapunov times, which are defined as the inverse of Lyapunov exponents, may be related to dynamic timescales associated with small bodies.
The relationship between LT and the dynamical lifetime of asteroids is known and was described in details by \cite{Soper1990} and \cite{Murison1994}. The dynamical lifetimes in these studies were defined differently and according to free interpretation as 'crossing time' and 'event time'.
It is characteristic that they mean the time interval after which the orbit is no longer stable. These times and LT values are often correlated, but sometimes a phenomenon called ``stable chaos'' occurs when a short Lyapunov time corresponds to a long time of orbital stability. As mentioned earlier, similar cases have been observed in our previous results. This probably means that a few objects with a large $\tau$ are in a different dynamical regime than the others.
The conclusive discussion on these relationships was presented by \cite{Morbidelli1996}. The authors suggest that the relationship between ``instability time'' and LT may be exponential or polynomial, depending on the particular dynamical regime. However, they suggest caution in treating such dependences as universal laws.

In most cases,  Lyapunov indicators (LIs) are estimated numerically at finite intervals of time. The proper selection of parameters for such calculations is crucial, as it can lead to different results that are not always consistent with the results of a study with similar or slightly altered initial assumptions \citep{Tancredi2001}. Taking into account these limitations, we decided to estimate the Lyapunov characteristic indicators (LCIs) and LTs of our group of retrograde asteroids to produce a final estimation of the chaotic properties of their orbits. Our hypothesis that the quantified chaos indicators are directly related to the previously studied dynamical lifetimes \citep{Kankiewicz2017} has also been verified.

There are several methods to estimate quantitatively chaos in orbital motion, such as MEGNO\footnote{Mean Exponential Growth factor of Nearby Orbits}, SALI\footnote{Smaller Alignment Index}, FLI\footnote{Fast Lyapunov Indicator}, and RLI\footnote{Relative Lyapunov Indicator}. Some of these are based on modified LIs, but there are many practical differences between these methods in particular applications. Many experiments and tests of the reliability of these indicators were performed by \cite{Maffione2011}. This study confirms the applicability of LI-based methods and gives some recommendations for the use of particular variants.

Because of its simplicity and popularity, the basic method of calculating LCI was used in this paper. This choice also allows us to compare our results with other studies. However, there are known limitations \citep{Tancredi2001} of this method that can be taken into account in the dynamics of small bodies.

The application of the two-particle method of LCI estimation is generally recommended in cases of strong chaoticity. This fact influenced the method we chose. Because previously \citep{Kankiewicz2017} we obtained relatively short dynamical lifetimes of retrograde asteroids (on the order of millions of years), these orbits can be suspected to be more chaotic than most small bodies. Our procedure for the numerical calculation of LCI was based on methods originally introduced by \cite{Benettin1976}, adapted and described in detail by \cite{Murray2000} and \cite{Sprott2003}.

As mentioned above, incorrectly set parameters used in the LCI and LT estimation may affect the results. Performing numerical experiments with different sets of these parameters helps to select their appropriate values. \citep{Tancredi2001}. Crucial parameters include the renormalization time and the initial separation of orbits in six-dimensional space (orbital elements or coordinates with velocities). However, other parameters, such as the step size or the total time of integration, may also influence the estimation of the LCI. Very often, the integration time may be too short for the LCI determination procedure to achieve convergence. On the other hand, the use of a billion-year timescales is risky if we take into account the evolutionary status of the asteroid families, planets, and the Solar System. 

In this study, the LCIs of 31 retrograde asteroids were determined for a finite interval of time of $10^7$ years. After tests with various values, the renormalization time was tuned to $t = 1000$ y. For most objects, this time was sufficient for detecting the chaotic divergence of the trajectories.
The optimal value of the initial separation parameter was $d_0 = 10^{-8}$. Since no significant variations between the final LCIs and LTs were observed, we assumed that the values of the parameters mentioned above could produce acceptable results. For uniformity, we applied these parameters in all cases and to all objects.

\subsection{Dynamical models}\label{model_section}
We also examined the issue of the dynamical model. Owing to the potential presence of non-gravitational effects typical of the asteroid (e.g. the Yar\-kov\-sky effect), we decided to include them in a similar form as in our previous studies \citep{Kankiewicz2018}. The influence of this effect was investigated. Later, in the next step, we implemented cometary effects.

In our previous estimation of LTs, a simple gravitational model with eight perturbing planets (excluding non-gravitational forces) was used. In these integrations, we used the original SWIFT package, precisely the RMVS method \citep{Duncan1998}. To obtain further results, the dynamical model presented in this paper was improved to accommodate the inclusion of non-gravitational effects that are present in both the dynamics of asteroids (Yar\-kov\-sky effect) and comets, as non-gravitational accelerations associated with cometary activity. 

To reproduce Yarkovsky forces in numerical integration, a complete set of physical properties was necessary for each asteroid.
This set included the following parameters: radius, albedo, bulk/surface density, thermal conductivity, thermal capacity, IR emissivity, rotation period, and spin orientation. In most cases, many of these values are unknown. Therefore, we found typical values presented in literature; this selection is described in detail by \cite{Kankiewicz2017}. In eight particular cases, asteroid diameters were available in literature and databases. These values are additionally denoted in the Table \ref{lt-table}.
In other cases, the diameter was estimated by using the albedo $p_v$ and absolute magnitude $H$ from the following formula \citep{Fowler1992}:
\begin{equation}
\label{diameter_formula}
D[km]=\frac{1329}{\sqrt{(p_v)}}10^{-\frac{H}{5}}.
\end{equation}

The problem of choosing the right physical parameters for estimating asteroid diameters is not trivial, especially when albedo is not determined. Usually in such cases the average value, characteristic for a given class of objects, is considered. The average albedo, in case of unavailability, has been assumed from \cite{Bauer2013} and \cite{Nugent2012a} (for NEA) works. Another comprehensive source of TNO and Centaur albedo is the database on Johnston's archive\footnote{\href{http://www.johnstonsarchive.net}{http://www.johnstonsarchive.net}}, where the approximate diameters of small bodies based on the averaged albedo were also determined. Further details can be found at \citep{Johnston2016}. The average albedo that we assumed on the basis of the available sources was therefore varied because the sample included TNOs, Centaurs, and NEA. Depending on the source, the mean albedo of Centaurs and scattered-disc objects (SDO) is 0.08 \citep{Bauer2013} and Johnston's archive gives a very similar value, that is 0.09 for Centaurs and TNO. This confirms that the data from both sources are consistent and can be used before individual results are available. The assumed values for bulk and surface density as 1125 $kg$ $m^{-3}$ for Centaurs and TNOs, and 1200 $kg$ $m^{-3}$ for retrograde NEAs. For the remaining parameters, we used the following values: for Centaurs and TNOs, $\kappa=0.006$ $W$ $K^{-1}m^{-1}$, $C=760$ $J$ $kg^{-1} K^{-1}$, $\epsilon=0.9$, for the mean period (used if rotation period was undetermined), $P=8.4$ h; and for the retrograde NEA, $\kappa=0.08$ $W$ $K^{-1}m^{-1}$, $C=500$ $J$ $kg^{-1} K^{-1}$, $\epsilon=0.9$, $P=f(R)$. 

When modelling the Yarkovsky effect, all physical parameters are rarely available for the studied objects.
Like most authors, we had to collect this information from many sources. The albedo of several objects was determined and available in the JPL Small-Body Database (JPL SBDB), but for the others, we had to take values according to \cite{Bauer2013} for Centaurs and NEA from \cite{Nugent2012}. 
For some objects, there was a rotation period available, and then we used real values. Therefore, we assume that for some individual objects, we chose the rotation parameters very close to the real values. The $\kappa$ and $\epsilon$ parameters mentioned above should be treated as common in use \citep{Broz2006}. We adopted NEA densities in the same way as \cite{Carruba2014} and other thermal parameters for Centaurs according to \cite{Guilbert2011}. Typical rotation periods and densities for TNO/Centaurs are taken from \cite{Sheppard2008}.

Unfortunately, pole coordinates were not available for all 31 asteroids. Since the spin axis orientation is crucial in estimating the Yarkovsky effect, we decided to analyse the possible extreme values of Yarkovsky drift by setting the obliquities ($\gamma$) to 0 and 180 degrees.
Assuming maximum or minimum impact of the Yarkovsky effect, depending on the obliquity value, is a typical strategy used in the simulations of studying the possible impact of this effect \citep{Nugent2012}.

The tool used for these numerical integrations was a modified version of SWIFT, known as the SWIFT\_RMVSY package \citep{Broz2011}, which was designed for simulations with Yarkovsky effect. Exceptionally, we procured an additional simulation with the model of the cometary forces for 333P (described in Section 3). This model took into account cometary, non-gravitational acceleration components, based on $A_1$, $A_2$, and $A_3$ parameters determined from available observations. In this particular case, MERCURY software \citep{Chambers2000}, which includes the Hybrid integrator and the possibility of applying user-defined perturbing forces, was adapted to this simulation. The last issue is to select the right amount of test particles, usually called ``clones''. Since a single object (the nominal orbital solution) does not reflect all possible orbital motion variants, we used various approaches to investigate probable scenarios and possible error propagation.

In the part of calculations in which we studied the Yarkovsky effect for retrograde asteroids and its influence on LIs, propagation of the separation vector of two particles in different directions was studied. Since the direction of the largest propagation of this vector was searched iteratively, up to 10 000 variants of this propagation were studied, integrating in practice almost as many newly cloned particles for each asteroid. As written above, this approach allows us to iteratively find the most extreme chaotic behaviour, which concerns variants only with gravitational forces and, additionally, when the Yarkovsky effect is strongest (obliquities of 0 and 180 degrees).

In the further part of the study, in which the cometary effects are examined, we had two parameters in the model, describing accelerations. The procedure was slightly different in this part: first, the extreme possible range of cometary accelerations $A_1$ and $A_2$ were studied. We examined LTs of 100 clones lying close to the nominal solution, distributed uniformly over the full error range of $A_1$ and $A_2$ values, for comets 333P and 161P.

Such a solution covers the probability space very well, but it can be slightly improved when the distribution of the most probable solution, the line of variation (LOV), is known. Therefore, we examined another option with the most statistically significant values, including other 100 clones lying on LOV. 
The LOV concept was proposed and described by \cite{Milani1999} and is widely used, especially by OrbFit software \citep{Orbfit} users. The LOV is usually obtained by fixing the value of all orbital elements, and it is a part of a curved line in the initial conditions space, corresponding to the statistical $3\sigma$ criterion. The LOV is used to compute multiple solutions that are useful in the orbit determination and their propagation. For each comet, we generated 100 clones uniformly distributed along the LOV and then calculated their LT values. The results are discussed in detail in Section \ref{cometary_section}.

So far, close approaches and resonances are considered to be the most critical factors influencing the chaos. Therefore, the question arises as to how much stronger these effects are than the non-gravitational effects and how accurate the dynamical model must be to compare them properly. In order to judge how subtle perturbations are needed in the initial dynamical model, we conducted numerical tests with models of varying degrees of complexity. These tests were done with REBOUND software \citep{Rein2012}, using numerical methods with the same functionality as described before \citep{Rein2015}, \citep{Rein2019}.

\begin{figure*}
\includegraphics[width=\textwidth]{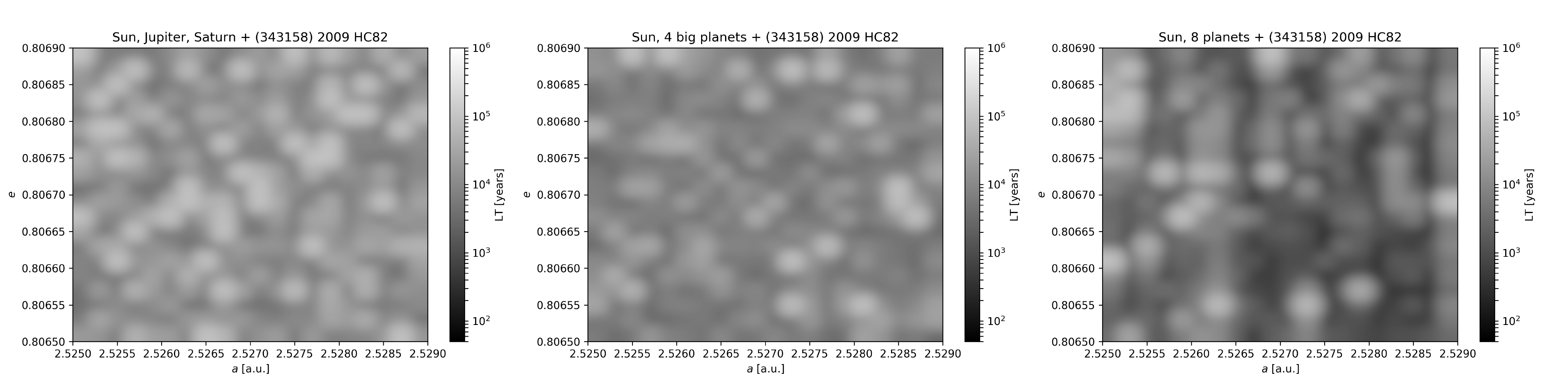}
\includegraphics[width=\textwidth]{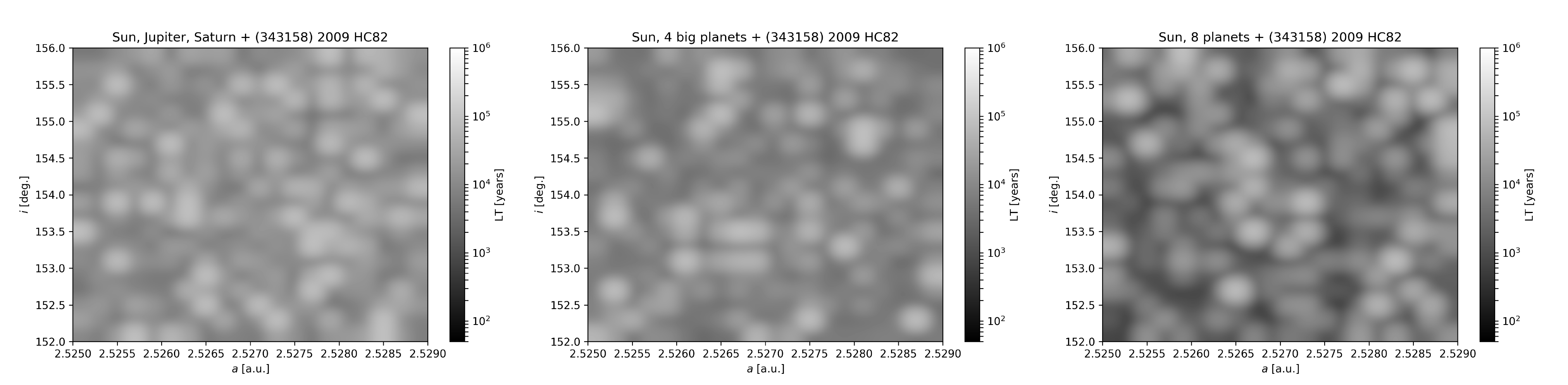}
\caption{Distribution of LT values in $a$,$e$ (top) and $a$,$i$ (bottom) space of the asteroid (343158) 2009 HC82 as the results of application of three different dynamical models (400 clones for each case)\label{spectrum_lt_343158}.}
\end{figure*}

\begin{figure*}
\includegraphics[width=\textwidth]{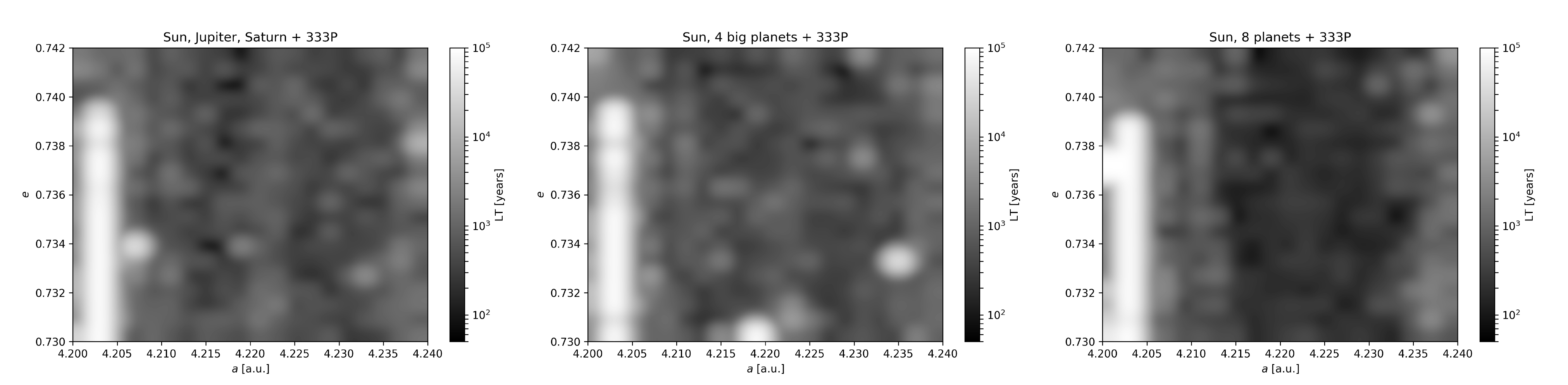}
\includegraphics[width=\textwidth]{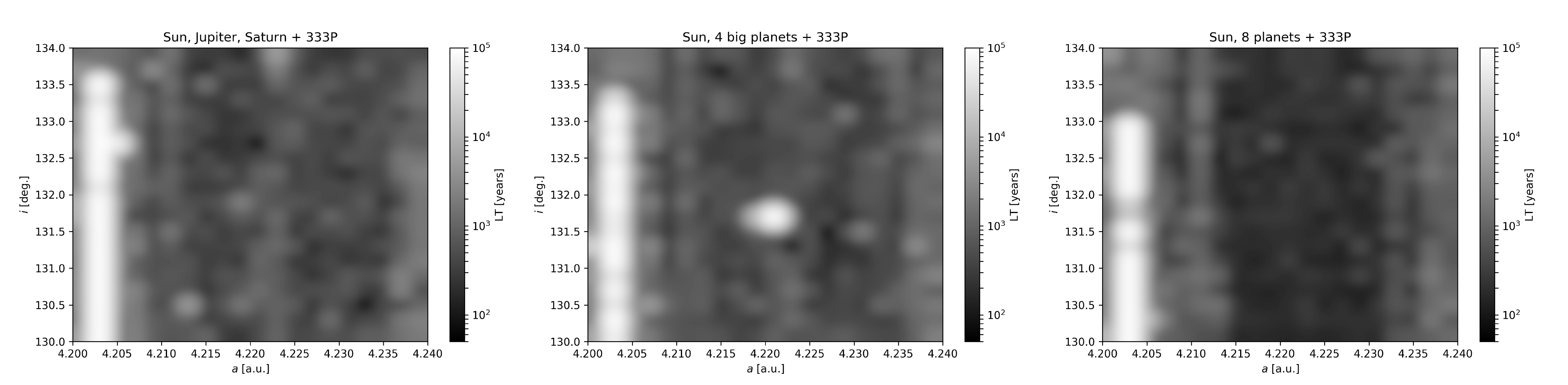}
\caption{Distribution of LT values in $a$,$e$ (top) and $a$,$i$ (bottom) space of the comet 333P as the results of application of three different dynamical models (400 clones for each case) \label{spectrum_lt_333}.}
\end{figure*}

\begin{figure*}
\includegraphics[width=\textwidth]{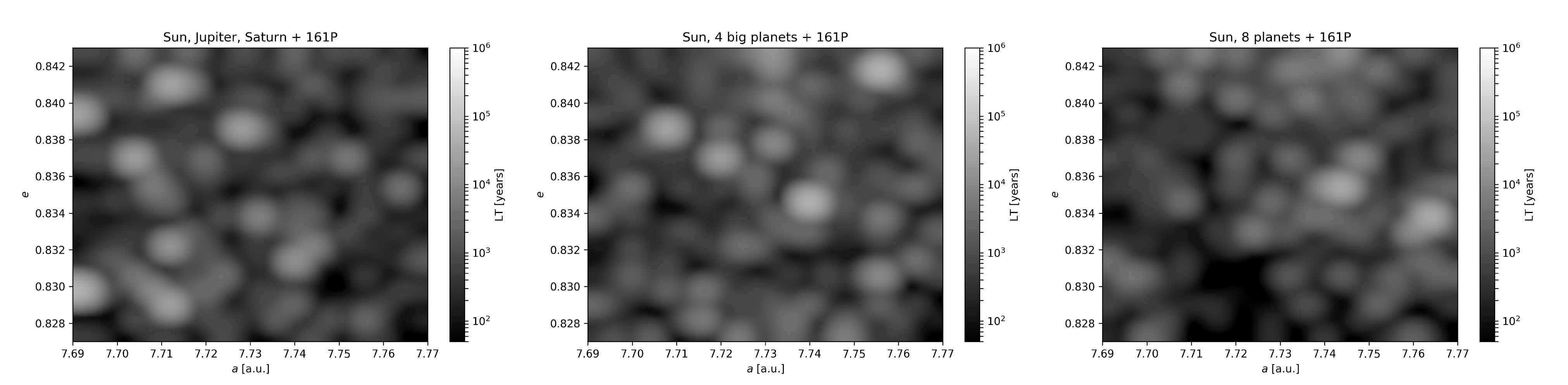}
\includegraphics[width=\textwidth]{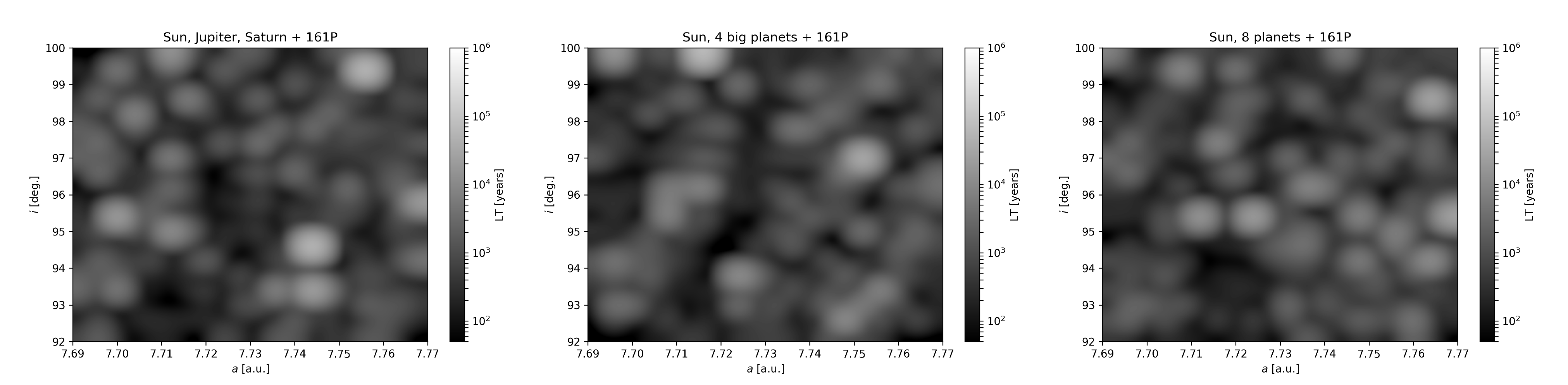}
\caption{Distribution of LT values in $a$,$e$ (top) and $a$,$i$ (bottom) space of the comet 161P as the results of application of three different dynamical models (400 clones for each case) \label{spectrum_lt_161}.}
\end{figure*}

The first, simplest model assumed only perturbations of a small body by Jupiter and Saturn. In the second model, we included four large planets. In the third, the most complex model we included eight planets. In each integration variant, 400 clones evenly distributed near the nominal solution were included. Solutions were shown in the space of elements $a,e$ and $a,i$. The results of such tests for one NEA and two comets are shown in the Figures \ref{spectrum_lt_343158}, \ref{spectrum_lt_333}, and \ref{spectrum_lt_161}. 

Figure \ref{spectrum_lt_343158} shows LT distribution near the nominal orbit of the asteroid (343158) 2009 HC82. Simulations in the range of $a,e$ and $a,i$ were performed for 400 clones and with three variants of planetary perturbations. As this asteroid is approaching the terrestrial planets, an additional chaotic factor is seen in the graphs on the left as darker, more chaotic areas.

A similar result can be observed for the comet 333P, as shown in Figure \ref{spectrum_lt_333}. This comet frequently approaches to Earth. Moreover, in this case, a more stable (lighter) zone is visible on the left side of all the plots. The presence of this zone is a result of perturbations caused by Jupiter.
The second comet shown in this figure, 161P, is an object in a near-polar orbit, and this leads to the consequences visible in Fig. \ref{spectrum_lt_161}. The number of close approaches is much smaller, so the complication of the dynamical model by adding more perturbers does not affect the results as much as in the previous cases.

Although each of the mentioned objects requires an individual approach, all LTs are generally the shortest if a model with eight planets is used. To get the complete picture of chaotic behaviour, this model should be applied. The only question that remains is whether non-gravitational forces can significantly change this image. We attempt to answer this question in the next sections.

\section{Results}

\subsection{Influence of Yarkovsky drift}
Using a typical strategy to estimate the maximum and minimum Yarkovsky drift (by modelling prograde and retrograde rotation), we investigated the effect of this force on the stability of the retrograde asteroids. Although for large asteroids this drift does not seem to matter, we found some regularities, partly confirming our previous results. Prograde rotation and, consequently, maximum drift in the semimajor axis seem to stabilize the orbits more often. In these cases, the LTs are longer (Tab. \ref{lt-table}).

In a few cases, Yarkovsky forces significantly raised the value of LTs, these cases are asteroids (342842) 2008 YB3 and 2013 BL76, the first of which is a relatively large asteroid ($H=9.3$ mag.). Such results occur under the $\gamma=180^{\circ}$ model, which produces highly stable solutions (LTs of $\sim 10^8$ years or more). Conversely, an interesting solution is available for the only known NEA: (343158) 2009 HC82. If we assume maximum drift at $\gamma=0^{\circ}$, Yarkovsky effect can significantly improve the stability of this orbit in the sense of Lyapunov. Object (514107) 2015 BZ509 deserves a separate description. It is an example of an object in 1:1 resonance and co-orbital motion with Jupiter. It is now the object that remains in the strongest retrograde resonance of the known retrograde asteroids \citep{Li2019}. In this case, the LT increased three orders of magnitude for both $\gamma=0^{\circ}$ and $\gamma=180^{\circ}$. This can be temporarily interpreted as stabilizing the orbit under the influence of possible non-gravitational effects. The asteroid is a unique example and is described in more detail by \cite{Wiegert2017}. There are also some indications that this object has an interstellar origin \citep{Namouni2018}.
\begin{table*}
\footnotesize\addtolength{\tabcolsep}{-3.5pt}
\begin{tabular}{|c|c|c|c|c|c|c|c|c|c|c|c|}
\hline
N & Name & $i$ [deg] & $a$ [au] & $e$ &          $H$[mag]         &  $D$[km] &      $LT_{GRAV}$[y]&  $LT_{\gamma=0^{\circ}}$ [y] & $LT_{\gamma=180^{\circ}}$[y] & $LT_{\gamma=0^{\circ}}$ & $LT_{\gamma=180^{\circ}}$\\        
 &  &  &  &   &  &  &   &  & & $-LT_{GRAV}$[y]  & $-LT_{GRAV}$[y]\\        
\hline
1&  (20461) Dioretsa          & 160.4&23.9&0.900&              13.8   &    8.17          &  5.77E4  &     9.75E4  & 3.78E4          &    3.98E4 & -1.99E4           \\
2&  (65407) 2002 RP120      & 119.0&53.9&0.954&              12.3   &      14.60$^1$         &  1E5     &     1.01E5  & 8.06E4          &    1E3 & -1.94E4              \\
3& {(330759) 2008 SO218}$^{\star}$& 170.3&8.1&0.564&               12.8   &11.80$^1$         &   1.76E5  & 1.32E5 & 1.7E5                &    -4.4E4 & -6E3              \\
4& {(336756) 2010 NV1}$^{\star}$  & 140.7&290.1&0.968&             10.4   &44.20$^1$         &   4.24E5 & 1.93E5 & 1.33E5                &    -2.31E5 & -2.91E5              \\
5& (342842) 2008 YB3        & 105.1&11.6&0.441&              9.3    &      67.10$^1$         &  $\ge$1.64E6  &   4.18E5  & $\ge$4.06E8   &   -1.22E6 & 4.04E8              \\
6& (343158) 2009 HC82       & 154.4&2.5&0.807&               16.2   &      2.04          &  1.71E5  &     $\ge$6.79E8  & 4.45E5     &    6.79E8 & 2.74E5              \\
7& (434620) 2005 VD         & 172.9&6.7&0.252&               14.3   &      6.49          &  $\ge$2.55E5  &   1.16E5  & 8.76E4       &    -1.39E5 & -1.67E5              \\
8& (459870) 2014 AT28       & 165.6&10.9&0.404&              12     &      18.71         &  1.27E5  &     1.28E5  & 8.17E4          &    1E3 & -4.53E4              \\
9& {1999 LE31}$^{\star}$  & 151.8&8.1&0.466&               12.38  &        16.80$^1$         &  1.04E5  &     7.62E4  & 6.42E4          &    -2.78E4 & -3.98E4              \\
10&        2000 HE46      & 158.5&23.6&0.899&              14.64  &        6.40$^1$          &  1.17E5   &     1.07E5  & 6.77E4         &    -1E4 & -4.93E4              \\
11&        2004 NN8       & 165.5&98.4&0.976&              15.25  &        4.19          &  5.41E4   &        4.67E4  & 1.06E5      &    -7.4E3 & 5.19E4              \\
12&        2005 SB223     & 91.3&29.4&0.905&               14.17  &        6.89          &  $\ge$5.65E5   &  $\ge$2.57E5  & 1.73E5  &    -3.08E5 & -3.92E5              \\
13&        2006 BZ8       & 165.3&9.6&0.803&               14.24  &        6.67          &  1.28E5   &     1.94E5  & 1.52E5         &    6.6E4 & 2.4E4              \\
14&        2008 KV42      & 103.4&41.7&0.493&              8.63   &        88.30         &  1.14E5   &        8.65E4  & 6.38E4      &    -2.75E4 & -5.02E4              \\
15&        2009 QY6       & 137.7&12.5&0.834&              14.74  &        5.30          &  1.49E5   &     2.05E5  & 8.52E4         &    5.6E4 & -6.38E4              \\
16&        2009 YS6       & 147.8&20.1&0.920&              14.27  &        6.58          &   1.67E5   &     2.57E5  & 6.97E4         &    9E4 & -9.73E4              \\
17&        2010 CG55      & 146.2&31.9&0.910&              14.19  &        6.82          &    2.11E5   &     4.39E5  & 1.08E5         &    2.28E5 & -1.03E5              \\
18&        2010 GW64      & 105.2&63.1&0.941&              14.84  &        6.42$^1$          &   5.38E4   &     8.75E5  & 3.68E4         &    8.21E5 & -1.7E4              \\
19&        2010 OR1       & 143.9&26.9&0.924&              16.14  &        3.25$^1$          &  $\ge$2.77E5   &  5.05E5  & 2.7E5        &    2.28E5 & -7E3              \\
20&        2011 MM4        &100.5&21.13&0.473&             9.3    &        63.70$^1$         &  1.33E5   &        1.08E5  & 1.32E5      &    -2.5E4 & -1E3              \\
21&        2012 HD2        &146.9&62.39&0.959&             15.3   &        4.09          &  $\ge$3.37E5   &  1.58E5 & 2.15E5        &    -1.79E5 & -1.22E5              \\
22&        2013 BL76       &98.61&1036&0.992&              10.8   &        32.51         &  1.39E5   &     7.51E4 & $\ge$1.27E9     &    -6.39E4 & 1.27E9              \\
23&        2013 LD16       &154.8&79.99&0.968&             16.1   &        2.83          &  2.04E5   &     2.44E5 & 2.51E5          &    4E4 & 4.7E4              \\
24& (468861) 2013 LU28       &125.4&175.5&0.950&             8      &      118.03        &  1.04E5   &     1.64E5 & 1.94E5          &    6E4 & 9E4              \\
25&        2013 NS11       &130.3&12.65&0.787&             13.6   &        15.24$^1$         &  1.18E5   &     2.05E5 & 2.74E5          &    8.7E4 & 1.56E5              \\
26 &{(471325)   2011 KT19$^{\star \star}$} &  110.1&35.6&0.330&7.2    &    170.60        & $\ge$3.22E8 & 2.18E5  & 5.36E4        &    -3.22E8 & -3.22E8              \\                         
27 &{(514107)   2015 BZ509$^{\star \star}$} & 163.0&5.1&0.380& 16     &    2.96          & 4E5    & $\ge$4.99E8  & $\ge$4.21E8    &    4.99E8 & 4.21E8              \\                                      
28 &{(518151)   2016 FH13$^{\star \star}$} &  93.6&24.5&0.614& 10.1   &    44.87         & 2.59E5 & 1.52E5  & 1.71E5             &    -1.07E5 & -8.8E4              \\                           
29 &{(523797)   2016 NM56$^{\star \star}$} &  144.0&72.7&0.855&11.4   &    24.66         & $\ge$2.71E8 & 4.73E5  & 8.09E5        &    -2.71E8 & -2.70E8              \\                         
30 &{(523800)  2017 KZ31$^{\star \star}$}  &  161.7&53.4&0.795&10.2   &    42.85         & 6.28E4 &  6.58E4 & 3.27E4             &    3E3 & -3.01E4              \\                              
31 & {2005 NP82$^{\star \star}$} &  130.5&5.9&0.478& 13.81  &              8.13          & 1.15E5 &  $\ge$1.4E9  & 1.94E5        &    1.4E9 & 7.9E4          \\                                                                                          
\hline
\multicolumn{12}{|l|}{$^1$ Diameter values taken from JPL SBDB (for other cases, approximate diameters were estimated with H and average albedo.)}\\
\hline
\end{tabular}
\caption{\label{lt-table} Lyapunov times obtained by three models: (1) gravitational, (2) with Yarkovsky effect, $\gamma=0^{\circ}$  and (3) with Yarkovsky effect, $\gamma=180^{\circ}$. In specific cases, owing to the lack of convergence in the LCI estimation procedure, some results are preceded by a $\ge$ sign, which can be interpreted as a time longer than or equal to the obtained value. Information updated since our previous studies \citep{Kankiewicz2018} is denoted with an asterisk. New objects in the list are denoted with a double asterisk.}
\end{table*}

When all the results are taken into general consideration, we can conclude that the model with $\gamma=0^{\circ}$ produces more stable orbits than the gravitational model, especially for small asteroids (Tab. \ref{lt-table}, two last columns). The described effect appears to be more visible for larger inclinations (closer to 180 degrees). If we look more closely at the changes in LT in the last two columns of Table \ref{lt-table}, we can see that the stabilizing effect (positive values) occurs more often than the destabilizing effect (negative values) for the model assuming $\gamma=0^{\circ}$. For the model with $\gamma=180^{\circ}$ (Table \ref{lt-table}, last column), we get less stable solutions. The statistical significance of these results for such a small population may be questioned, but after rejecting the largest asteroids (e.g. by estimating the approximate size from $H$), this effect becomes more visible.

The main factors affecting overall stability (in the Lyapunov sense) are regular close approaches to planets and resonances. In these approaches, we can see the causes of chaotic behaviour for individual cases in the orbital motion of asteroids.

Some studied objects showed a strong dependence of Yarkovsky forces on LT. This can be explained by the slow migration of the semimajor axis due to possible thermal accelerations, and consequently the transfer of asteroids to another dynamical regime. This may mean a different configuration of close approaches, which destabilize the orbit, but also migration to the zones accompanying specific mean motion resonances (MMR). In the case of retrograde asteroids, these are so-called retrograde MMR and abbreviated as RMMR \citep{Li2019}.

The LCI values, as mentioned earlier, are numerical LCE estimators at the finite interval of time $10^7$ y. This approach was necessary to compare the tested orbits between each other and to guarantee the internal consistency of the results. Unfortunately, retrograde orbits vary from a dynamic point of view, and it was  not easy to arbitrarily select the parameters, which in each case lead to a convergence of the solution (Table \ref{lt-table}). There is also no clear correlation of LT with previously calculated median lifetimes. We explain this paradox with different definitions of stability. If we compare selected, dynamically interesting asteroids, we notice why LT estimates differ for particular dynamical models and even identify potential causes of more or less stable solutions. Good candidates for such objects are (342842) 2008 YB3, 2008 KV42, (343158) 2009 HC82, and 2006 NP82.

\begin{figure}
\includegraphics[width=\columnwidth]{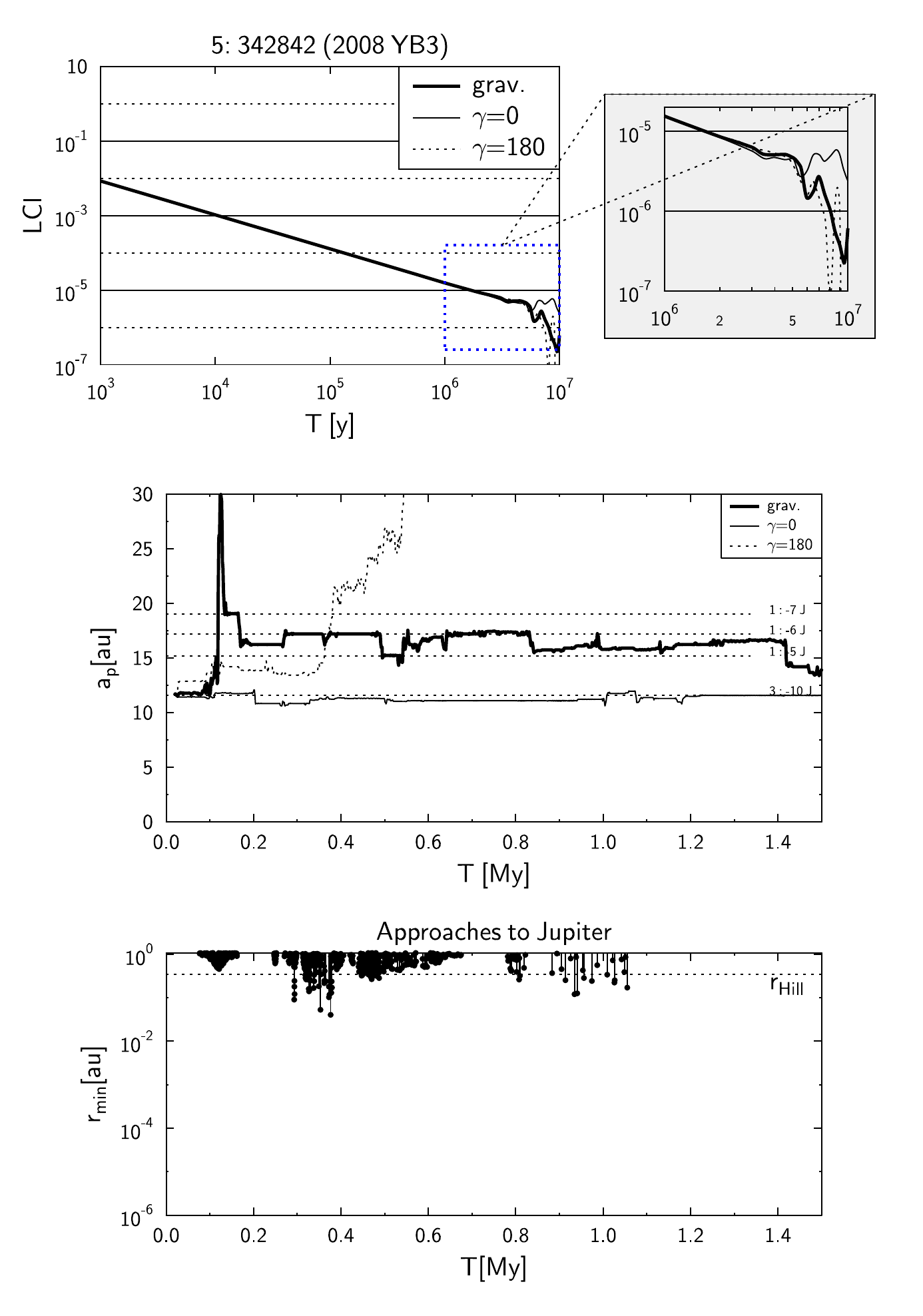}
\caption{Convergence of the LCI value for the asteroid (342842) (top), possible variants of the evolution of proper semimajor axis with the indicated RMMR with Jupiter (centre) and close approaches of the asteroid to Jupiter (bottom).\label{evolution_tp5}}
\end{figure}

The object (342842) is an example of the large retrograde Centaur, and it is also classified as Damocloid.
This object is in a retrograde resonance of 3:-10 with Jupiter. In the next few million years, the object will have many close encounters to Jupiter, and these perturbations will be the main source of migration to other dynamical regions. The behaviour of both the chaotic indicators and the potential factors causing their changes (RMMR and close approaches) can be followed in more detail. For (342842), this information is presented in Fig. \ref{evolution_tp5}. The top plot shows the convergence of LCIs, calculated with different dynamic models, which leads to the least stable solution for the $\gamma=0$ case. This corresponds to the case in the middle plot, where a semimajor axis remains close to the resonance 3:-10 J. It seems that this resonance causes the most chaotic behaviour. The condition to remain in this resonance is to maintain the value of $a$ after the cumulative number of close encounters with Jupiter (bottom plot), which depends on the assumed dynamical model. The influence of this particular, polar RMMR for  (342842) is also confirmed in \cite{Li2019}.

\begin{figure}
\includegraphics[width=\columnwidth]{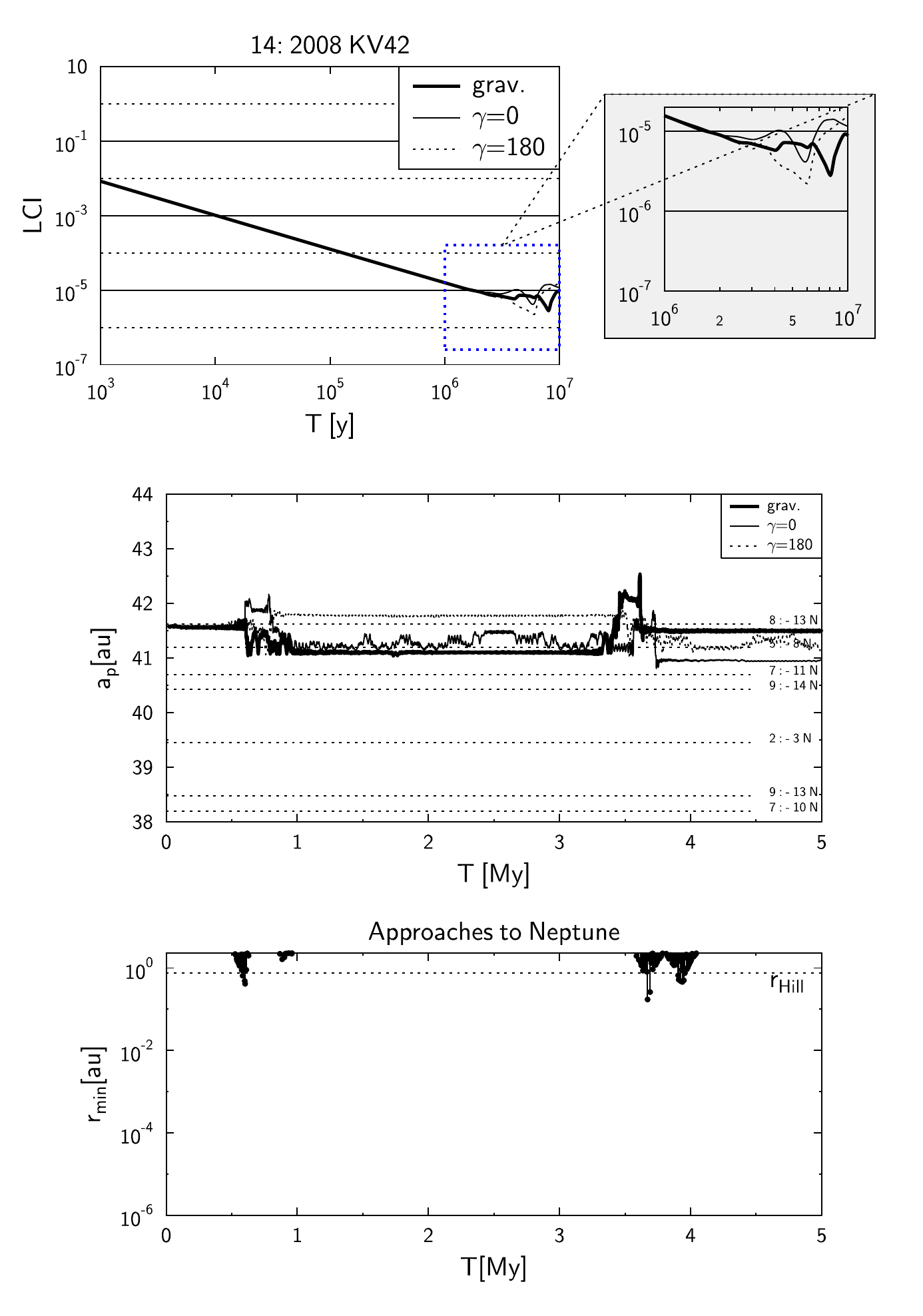}
\caption{Values of LCI for the asteroid 2008 KV42 (top), possible variants of the evolution of proper semimajor axis with the denoted RMMR with Neptune (centre) and close approaches of the asteroid to Neptune (bottom). \label{evolution_tp14}}
\end{figure}

The Transeptunian object, 2008 KV42, is known for its long median dynamical lifetime (\cite{Chen2016, Kankiewicz2017, Li2019}). The motion of this asteroid is dominated by 8:-13 N RMMR \citep{Li2019}. In this case, there are no approaches to Jupiter, but only regular series of close approaches to Neptune. It is worth noting that Neptune has the largest radius of Hill (0.77 au) among the planets, and therefore is an important source of regular perturbations. Detailed information on estimated LCI, RMMR, and close approaches to Neptune of 2008 KV42 is shown in Fig. \ref{evolution_tp14}.

\begin{figure}
\includegraphics[width=\columnwidth]{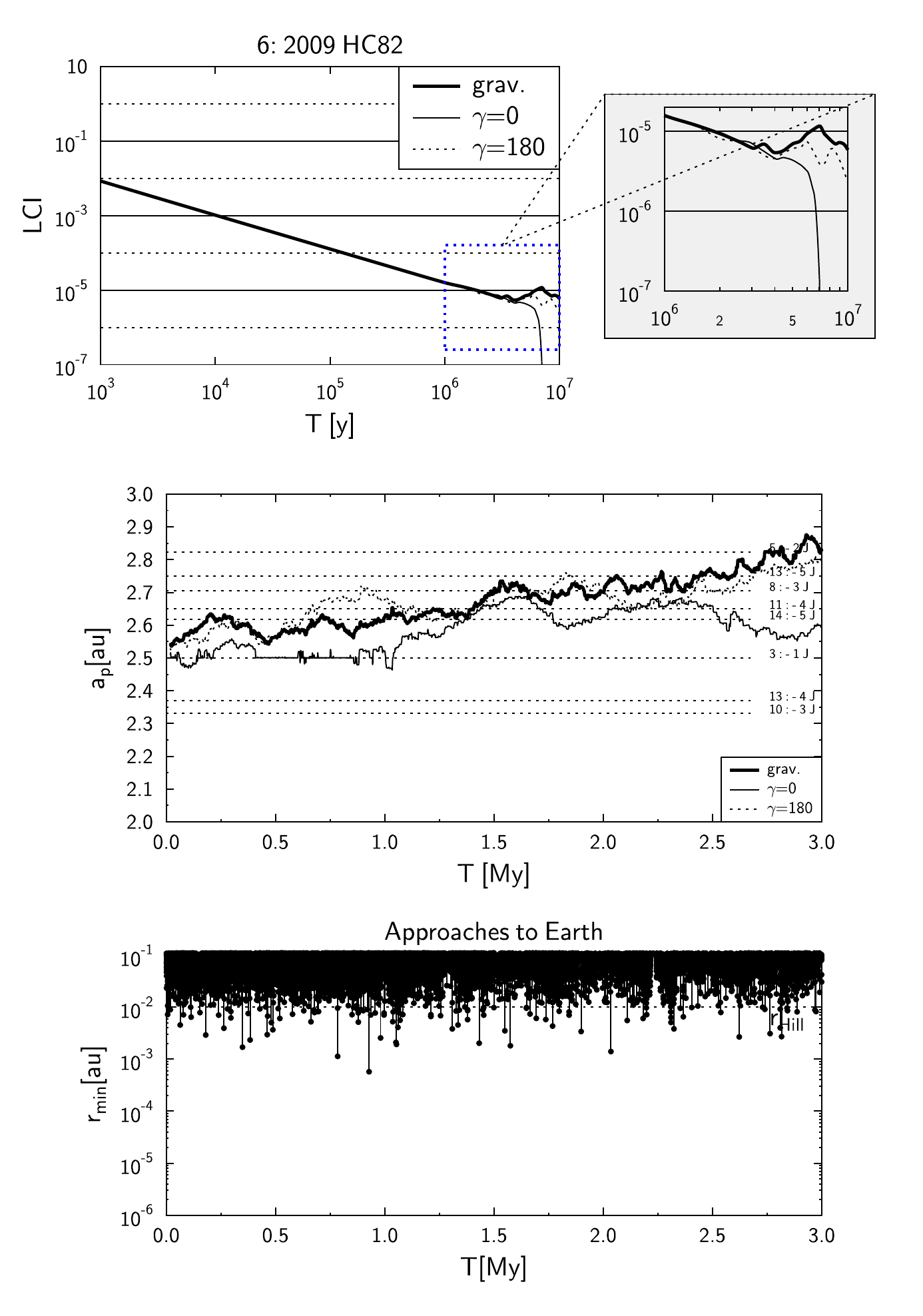}
\caption{Values of LCIs, semimajor axis, and close approaches to the Earth of the asteroid (343158) 2009 HC82.\label{evolution_tp6}}
\end{figure}

\begin{figure}
\includegraphics[width=\columnwidth]{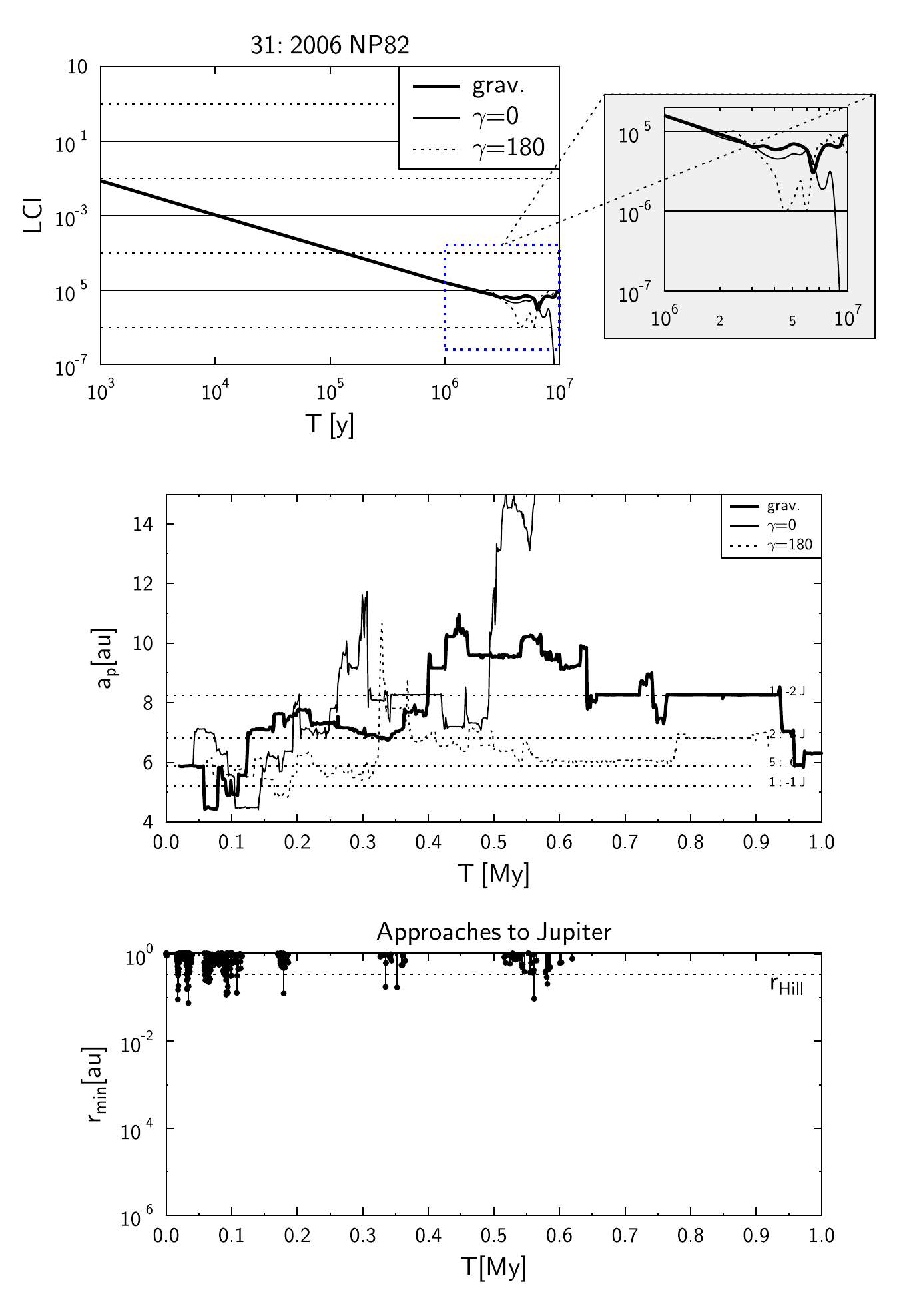}
\caption{Values of LCIs, semimajor axis, and close approaches to Jupiter of 2005 NP82.\label{evolution_tp31}}
\end{figure}

Similarly, the influence of selected resonances on stability can be observed for objects (343158) 2009 HC82 and 2005 NP82. (Figs. \ref{evolution_tp6}, \ref{evolution_tp31}). 
 Although the asteroid (343158) 2009 HC82 (Fig. \ref{evolution_tp6}) remains in resonances with Jupiter (currently 3:-1 J), it approaches mainly the Earth and Venus. For this asteroid, the solution with the model assuming $\gamma=0$ is less chaotic than others, because it corresponds to an area less covered by Jovian resonances. For 2005 NP82, the most stable solution in the sense of Lyapunov was obtained for the model with $\gamma=0$, where the asteroid was transferred to a distant orbit (Fig. \ref{evolution_tp31}).

More generally, there are possibly two reasons for the migration of an asteroid to an area dominated by a particular RMMR.
One of these is the presence of close encounters with planets causing characteristic jumps in the semimajor axis, and the second are slow changes in the semimajor axis due to the Yarkovsky effect. Whether a particular asteroid stays in the RMMR area for a longer time or leaves this area consequently determines its chaotic properties.

An additional factor is the relative strength of the mentioned RMMRs. Consider, as an example, several previously identified retrograde resonances \citep{Li2019}, which we also confirmed in our studies. One of the more remarkable examples is the near-polar resonance 7:-9 of (471325) 2011 KT19 with Neptune \citep{Morais2017}. Recent simulations of $10^5$ clones of this object, carried out by \cite{Namouni2020} have shown that the impact of this resonance on orbital evolution is negligible. This is not surprising because the strength of this resonance can be estimated as relatively weak. We can approximately estimate the relative strength of these resonances by assuming that the amplitude of dominant resonant term in a retrograde motion is proportional to $e^{p+q}$, where $e$ is eccentricity and $p$ and $q$ are integers characterizing the retrograde resonance $p$:$-q$ \citep{Morais2013}. This assumption fits best to near co-planar retrograde orbits, with inclinations close to 180 degrees. Nevertheless, we can roughly estimate which of the resonances are comparatively stronger. The relative force coefficients of RMMRs are proportional to the corresponding values for (342842) 2008 YB3 (in $3:-10$ J): $2.39 \times 10^{-5}$ , for 2008 KV42 (in $8:-13$ N) : $3.55 \times 10^{-7}$, and for (471325) 2011 KT19 (in 7:-9 N) : $1.98 \times 10^{-8}$. The unique exception in this comparison is the strongest RMMR (1:-1 J) of the object (514107) 2015 BZ509, which has an amplitude proportional to $1.44 \times 10^{-1}$. The other relative amplitudes are generally much smaller. For near-polar orbits, these amplitudes are more reduced. Therefore, it can be concluded that most of the mentioned resonances also have small widths. Consequently, it is expected that their long-term contribution to orbital evolution is not too extensive. In these cases, the time an object stays in the RMMR area during the evolution is short. This is also confirmed by the latest results of long-term simulations of retrograde objects \citep{Namouni2020}, especially in the context of the asteroid (471325) 2011 KT19, mentioned above.

\subsection{Influence of cometary non-gravitational forces}\label{cometary_section}
Significant non-gravitational effects on the motion of the studied objects take the form of thermal forces associated with Yarkovsky effect, as well as potential cometary effects. Since there is a strong hypothesis that a certain percentage of these small bodies show cometary activity, this scenario can be analysed for some individual objects. The best candidate is a retrograde asteroid reclassified as a comet: 333P (2007 VA85). For several years since its discovery, the small body 2007 VA85 has been known as a unique retrograde NEA \citep{KankiewiczWlodarczyk2010}. After observing a small tail in 2016, cometary activity was confirmed, prompting a significant revision to the dynamical model. With the arrival of new astrometric data regarding 333P, we decided to determine the non-gravitational acceleration parameters $A_1$, $A_2$ from the observations and then used them in our dynamical model (Tab. \ref{table-333p}). As a consequence, we had the opportunity to compare various non-gravitational effects and to examine their influence on the chaotic properties of the 333P orbit. 
\begin{table*}
\footnotesize\addtolength{\tabcolsep}{-4pt}
\caption{Retrograde comet 333P/LINEAR (2007 VA85): orbital elements (JD 2458600.5) and selected physical data\label{table-333p}.}
\begin{center}
\begin{tabular}{|c|c|c|c|c|c|c|c|}
\hline
Obj.   name     &        $a[au]$    &       $e$     &  $i_{2000}[deg]$ &     $\Omega_{2000}[deg]$    &       $\omega_{2000}[deg]$    &  $M[deg]$& no. of  \\
 & & & & & & & obs. used \\
\hline
333P/2007 VA85 & 4.22259& 0.735955 & 131.883 & 115.582 &  26.114 & 127.062 & 663 \\
1-$\sigma$ rms & 5.637E-06 &  2.660E-07 &  1.580E-05 &  1.844E-05 &  1.811E-05 &  2.106E-04 &  \\ 
\hline

\multicolumn{8}{|c|}{Physical data:}\\
\hline
\multicolumn{8}{|l|}{JPL NASA Classification: Jupiter-family comet [NEO]}\\
\multicolumn{8}{|l|}{Period: P = 21.04 h \citep{Hicks2016}; Cometary parameters, total magnitude, and its slope: M1=15.1, k1=7 }\\
\multicolumn{8}{|l|}{Cometary parameters NGR [$AU/d^{2}$]:}\\ 
\multicolumn{8}{|l|}{$A_1$=2.095E-10 $\pm$ 1.350E-10  (radial)}\\
\multicolumn{8}{|l|}{$A_2$=-1.906E-11 $\pm$ 1.181E-11 (transverse)}\\
\multicolumn{8}{|l|}{Possible Yarkovsky drift $\frac{da}{dt}$ based on formula of \cite{Nugent2012}: (1.63E-4 $\div$ 2.41E-4) $AU/My$}\\
\multicolumn{8}{|l|}{Earth MOID ({\it minimum orbit intersection distance}):  0.176206 $AU$}\\
\multicolumn{8}{|l|}{{\it Median dynamical lifetime}: $\tau \simeq \pm 0.5 \; My$.}\\
\hline
\end{tabular}
\end{center}
\end{table*}

Based on 663 published observations of the 333P LINEAR comet, we determined its orbit with the latest version of the OrbFit tool \citep{Orbfit}. This version (v.5.0.5) has the new error model based upon \cite{Chesley2010} and the debiasing and weighting scheme described in \cite{Farnocchia2015}. In this procedure, we used the DE431 version of the JPL planetary ephemerides and took into account 17 perturbing asteroids. We followed a similar approach as in the works of \cite{Farnocchia2013} and \cite{Wlodarczyk2015}. Finally, we received two solutions for non-gravitational parameters: one with three parameters $A_1$, $A_2$, $A_3$ and  a simpler solution with the parameters $A_1$ and $A_2$ (only radial and transversal acceleration). The second solution fit the orbit better and had much smaller errors, so we decided to adapt it to the model. The details are given in the Tab. \ref{table-333p}.

\begin{figure}
\includegraphics[width=\columnwidth]{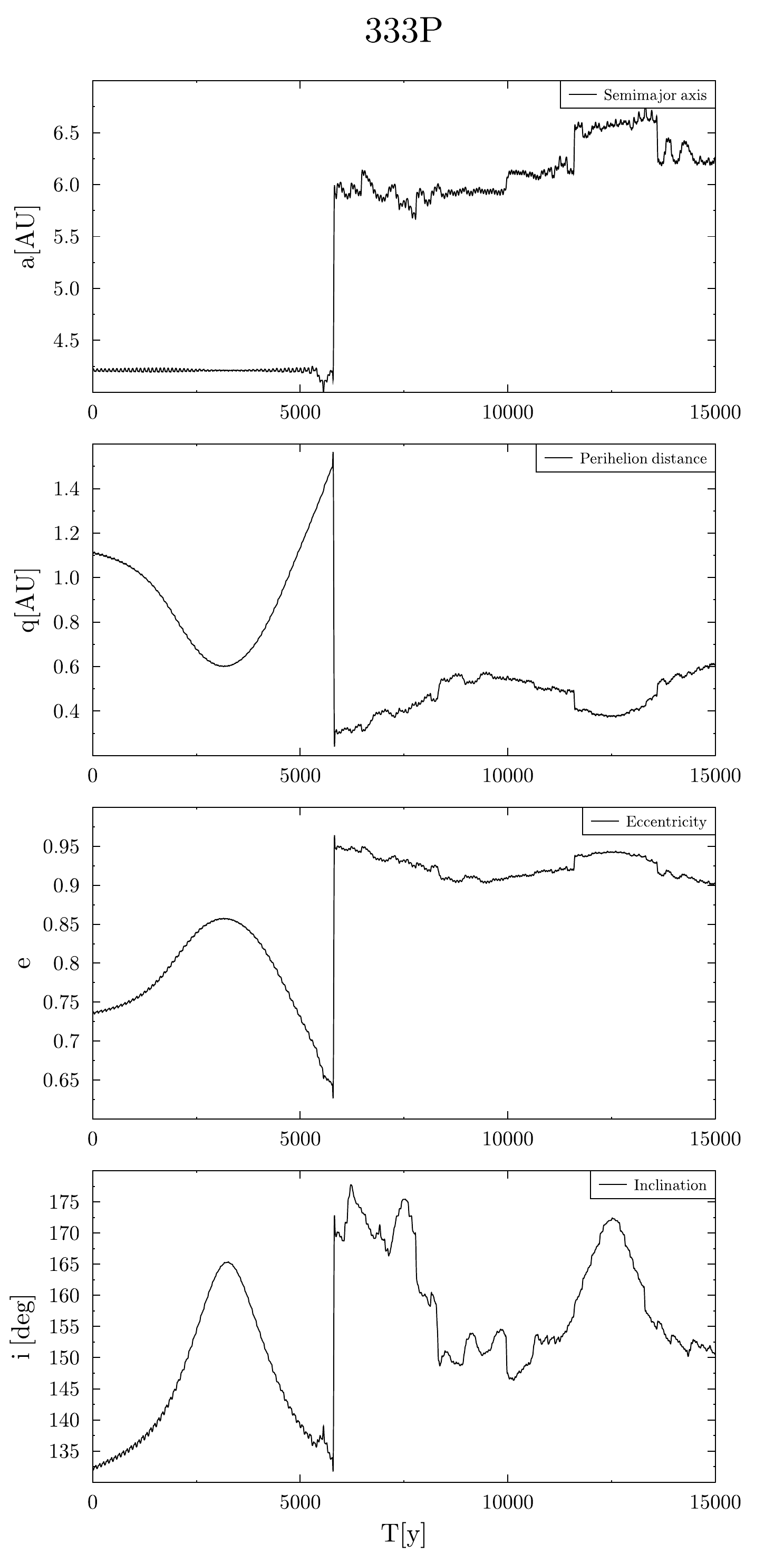}
\caption{Short-term dynamical evolution of semimajor axis, perihelion distance, eccentricity, and inclination of 333P.\label{evolution_333}}
\end{figure}

\begin{figure}
\includegraphics[width=\columnwidth]{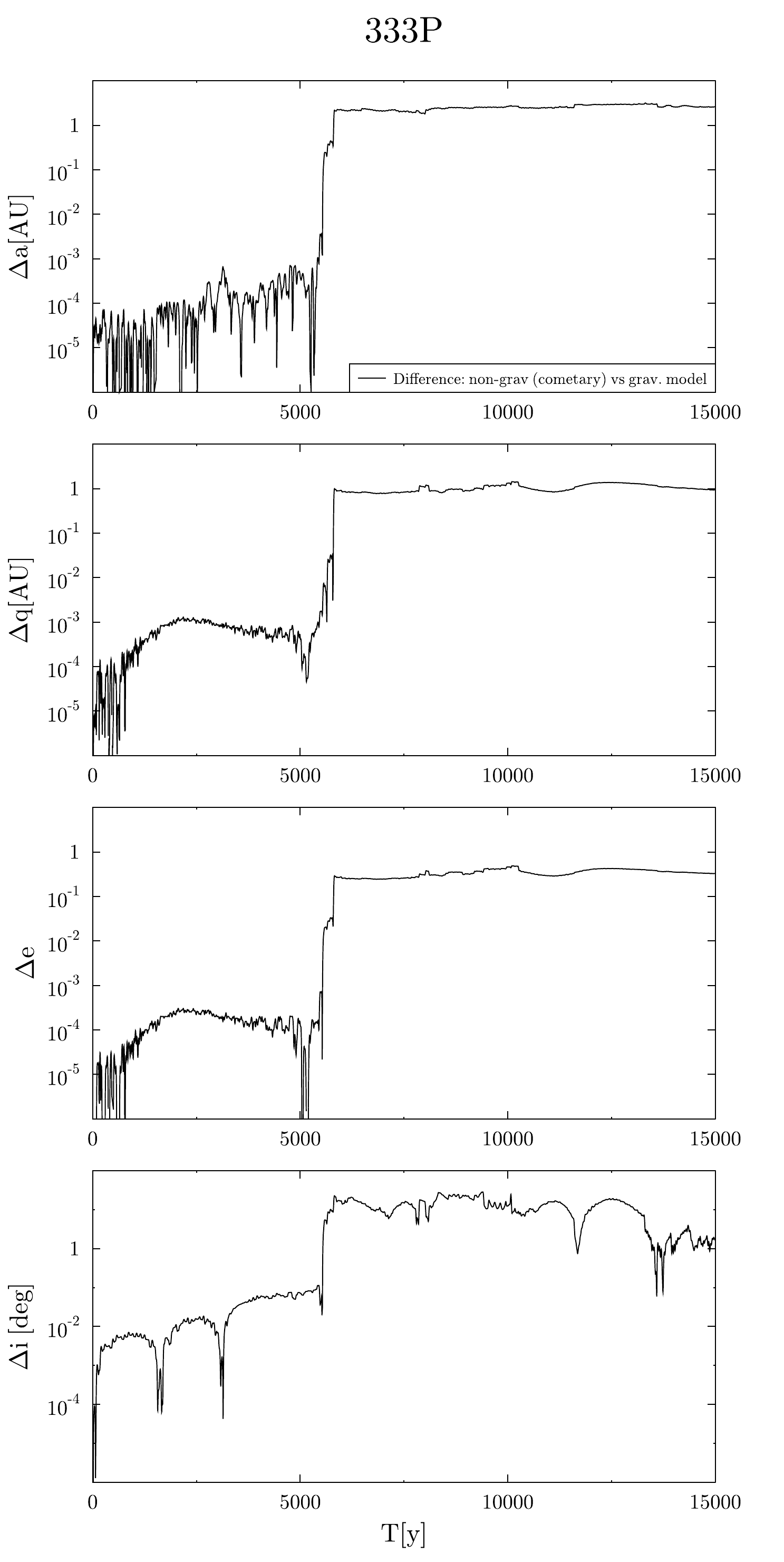}
\caption{Differences in orbital elements of 333P due to the application of the model with cometary forces in the numerical integration.\label{ngr}}
\end{figure}

\begin{figure}
\includegraphics[width=\columnwidth]{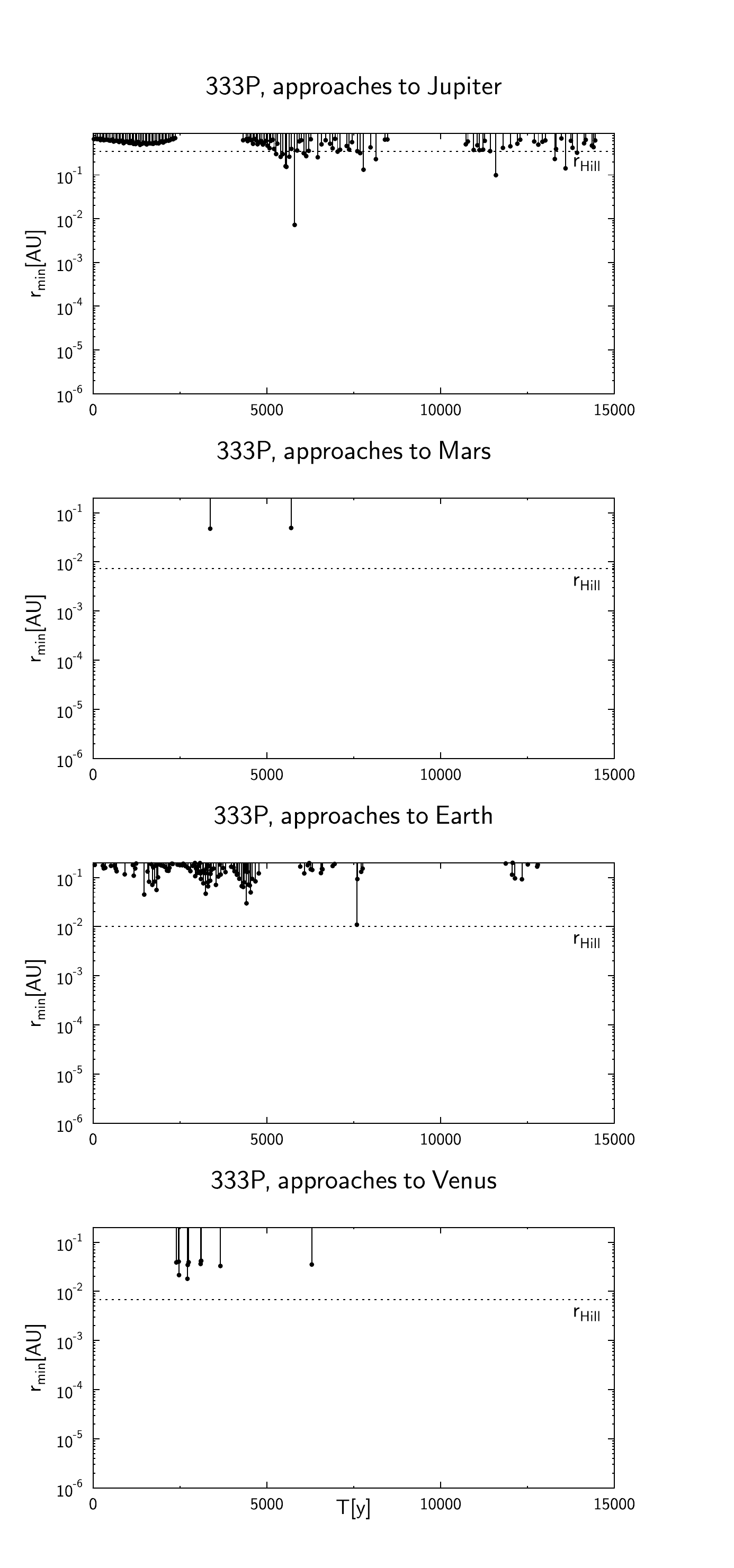}
\caption{Close approaches of 333P from Jupiter, Mars, Earth, and Venus. \label{appr_333P}}
\end{figure}

Despite the quite extensive observational data, the determined parameters of the cometary accelerations ($A_1$, $A_2$) are known with limited accuracy. Such big errors are caused by a relatively short observational arc, which was about ten years. At the moment, this was the only available data from the observations, so we decided to rely on these data to prepare a dynamical model. Using this approximate model of cometary activity, we decided to estimate its impact on the dynamical evolution and chaotic properties of the orbit. 
Such a simple model of cometary activity is not suitable for drawing conclusions about the long-term evolution and origin of a particular comet \citep{Wiegert1999}. This is because cometary activity over a long time is hardly predictable. However, to predict short-term evolution and the accompanying events is a good approximation. Especially when the LTs, as we will show below, are on the order of hundreds of years.
During the short-term evolution, 333P approaches the Earth and Jupiter (Fig. \ref{appr_333P}), which causes characteristic jumps in the semimajor axis (Fig. \ref{evolution_333});  in Figs. \ref{evolution_333}, \ref{ngr} and \ref{appr_333P}  we decided to show a part spanning 15 000
years in the future. The use of a more precise cometary model shows how differences in the semimajor axis, eccentricity, and inclination can propagate due to non-gravitational forces (Fig. \ref{ngr}). 
Practically, after several thousand years, these errors appear to have a significant influence on the results.

\begin{figure}
\includegraphics[width=\columnwidth]{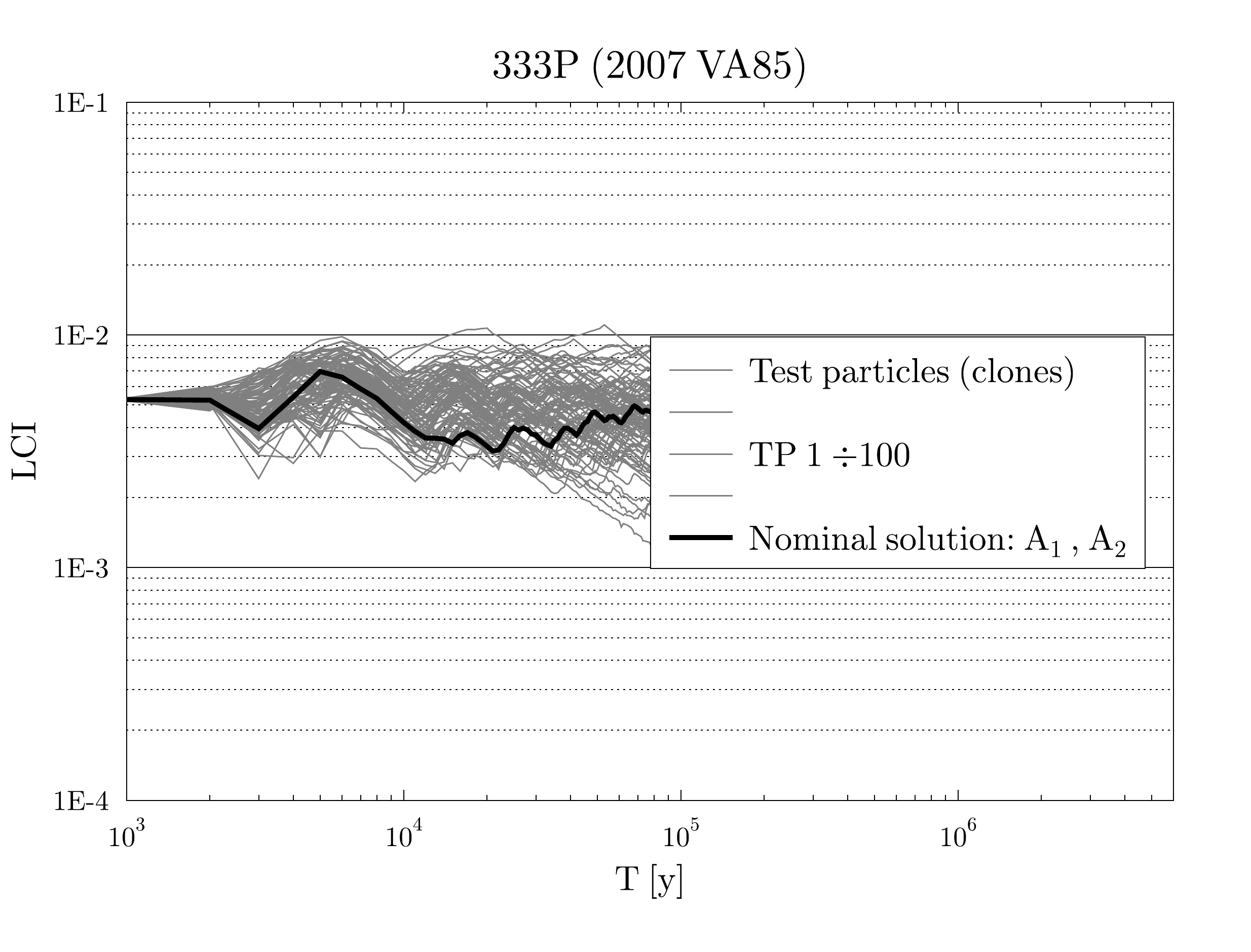}
\caption{Lyapunov indicators of comet 333P: dependence on non-gravitational effects.\label{333lci}}
\end{figure}

\begin{figure}
\includegraphics[width=\columnwidth]{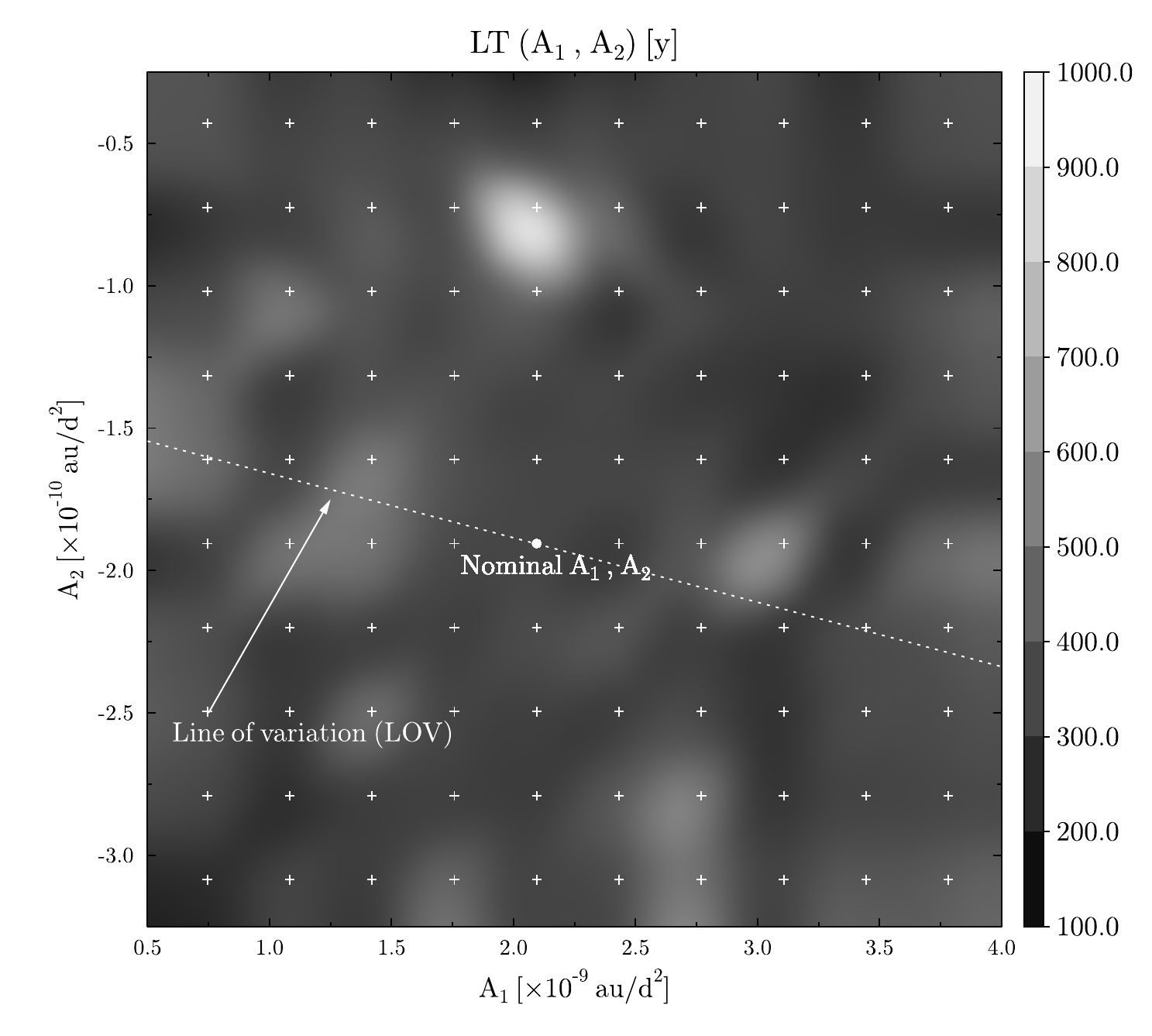}
\caption{Dependence of LT on non-gravitational acceleration parameters (model: $A_1$, $A_2)$ for comet 333P.\label{333lt}}
\end{figure}

\begin{figure}
\includegraphics[width=\columnwidth]{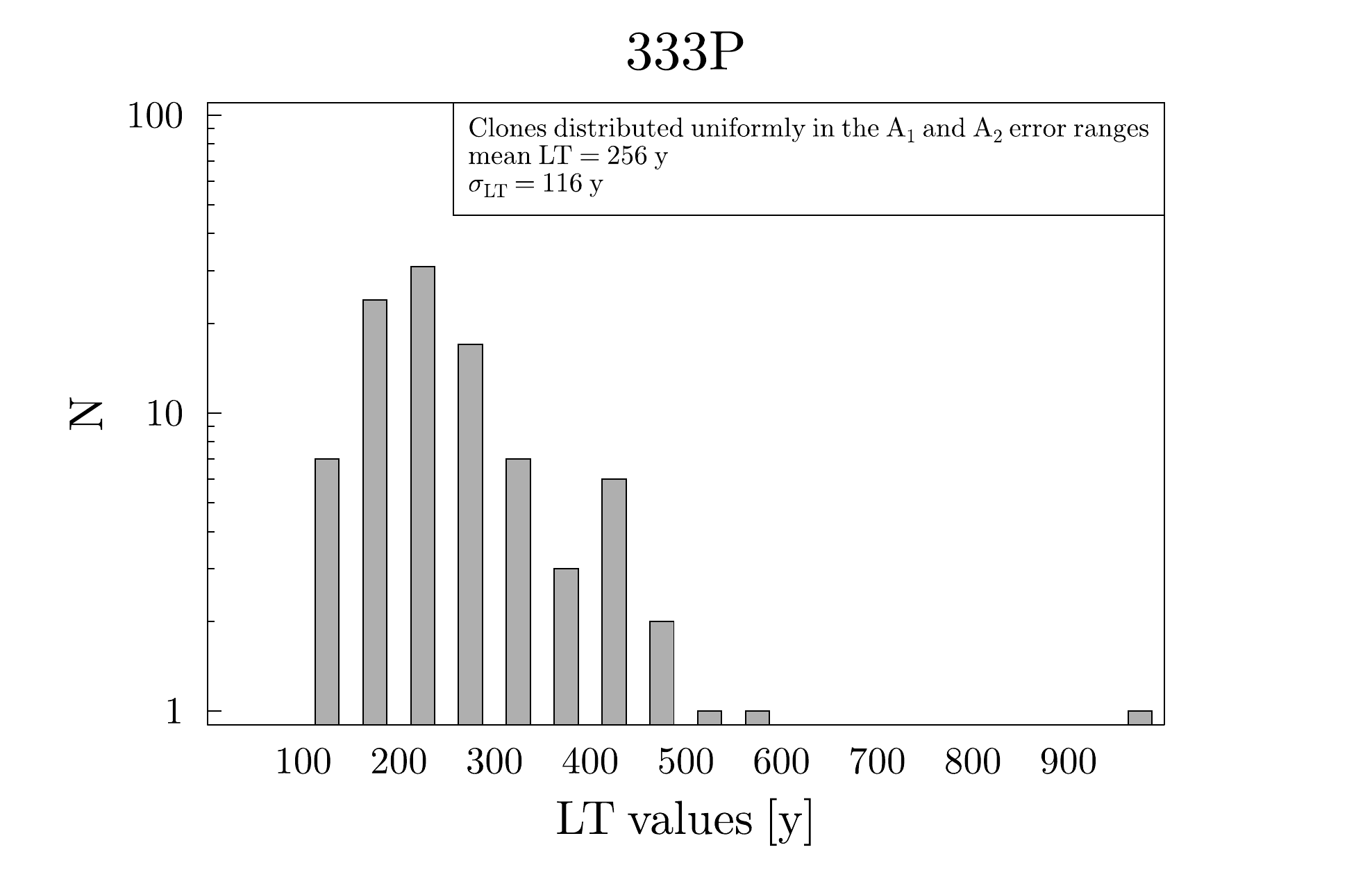}
\caption{Histogram of LT values for comet 333P. Result of the numerical integration of 100 clones, distributed in the error ranges of $A_1$, $A_2$. \label{333lt-hist-matrix}}
\end{figure}

Estimating the chaotic properties required a slightly different approach. The numerical calculation of the LCI/LT values using various models required a numerical integration with more extended time (min. about $10^5$ years). To investigate the $A_1$ and $A_2$ error propagation, we prepared data sets with adjacent nominal, maximum, and minimum values of these parameters. We also added a gravitational model in which the accelerations are equal to zero. As a result, we determined LCI values for 100 test particles. Fig. \ref{333lci} compares the behaviour of the LCI values in time. Figure \ref{333lt} demonstrates an approximate surface plot that shows the dependence between the possible coefficients of the cometary model and the obtained LTs. It also shows individual LT values calculated for each of the test particles. The histogram of these values is shown in Fig. \ref{333lt-hist-matrix}.

\begin{figure}
\includegraphics[width=\columnwidth]{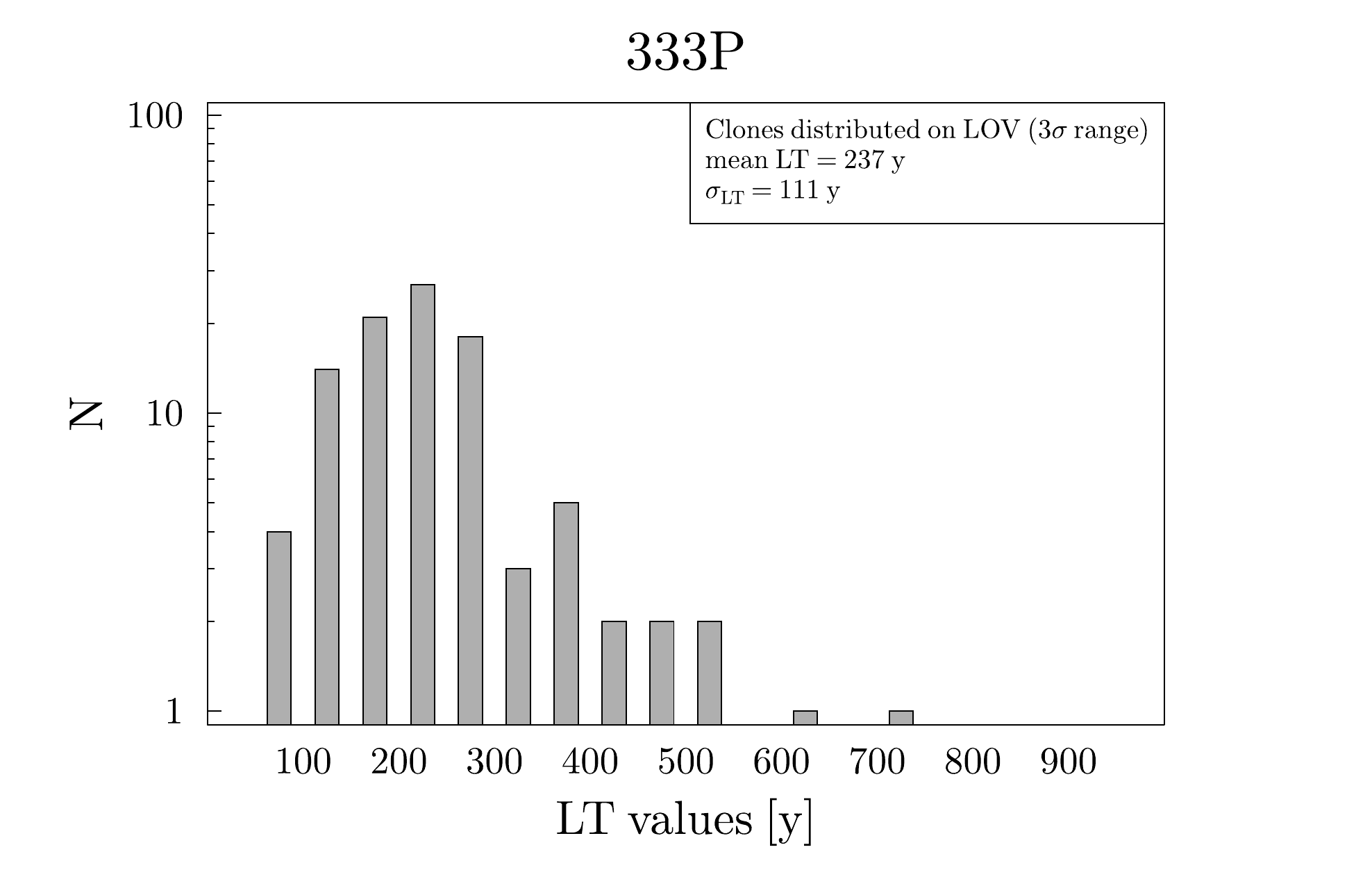}
\caption{Histogram of LT values for comet 333P. Result of the numerical integration of 100 clones, distributed on the LOV, close to the nominal solution. \label{333lt-hist-lov}}
\end{figure}

Alternatively, it is possible to investigate how such results behave in a more statistically significant region, known as the LOV, originally defined and described by \cite{Milani1999}. We selected  100 test particles close to the nominal solution laid on the LOV. The LT behaviour, in this case, is illustrated in the histogram (Fig. \ref{333lt-hist-lov}). The results do not vary on order of magnitude (hundreds of years) and they appear to be concentrated in the similar way around the nominal result. The LOV solutions appear to be comparable to those resulting from the previously used distribution of errors.
The final, possible LT values obtained, depending on the adopted non-gravitational parameters, range from 100 to almost 1000 years. The final LT values for 333P are of a similar order: $LT_{grav}=165$ y and $LT_{cometary}=255$ y. This particular near-Earth comet is noticeably less stable than the retrograde Centaurs and TNOs. However, it is apparent that cometary forces influence the chaotic behaviour of the orbit.

The above considerations were initially limited to one comet (333P). To better characterize the effect of non-gravitational effects on similar objects, we decided to look for comets on similar orbits with sufficient observational data for analysis. It is difficult to find such a candidate because although many comets have retrograde orbits, only a few of them are classified as NEO, and in general, very few have enough observations to estimate non-gravitational forces. Such a comet is 161P Hartley/IRAS, moving in retrograde, near-polar orbit. The orbital elements and the determined coefficients $A_1$ and $A_2$ for this comet, together with the corresponding errors, are given in Table \ref{table-161p}.

\begin{figure}
\includegraphics[width=\columnwidth]{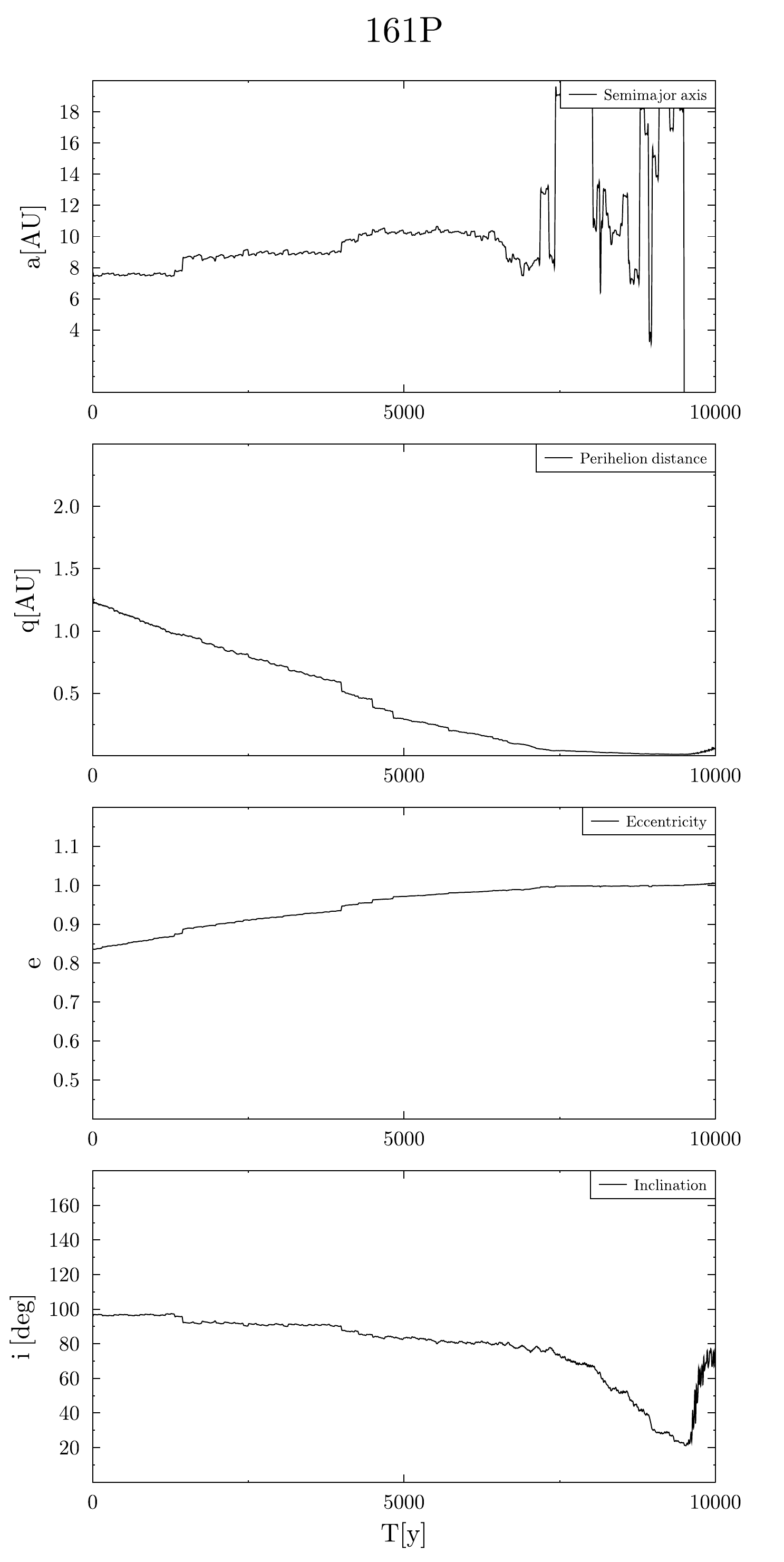}
\caption{ Short-term dynamical evolution of semimajor axis, perihelion distance, eccentricity, and inclination of 333P.\label{evolution_161}}
\end{figure}

\begin{figure}
\includegraphics[width=\columnwidth]{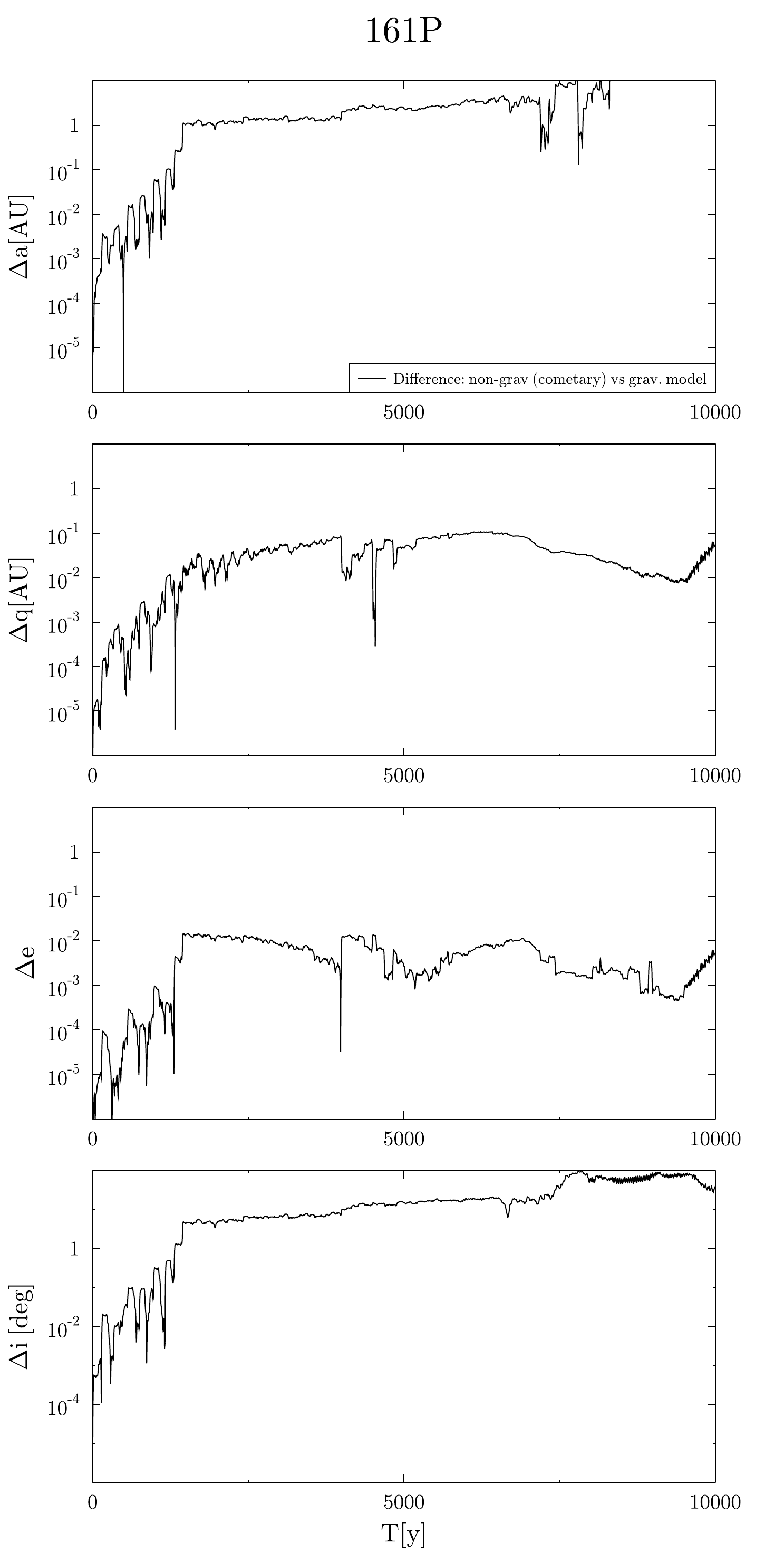}
\caption{Differences in orbital elements of 161P due to the application of model with cometary forces in the numerical integration.\label{ngr161}}
\end{figure}

\begin{figure}
\includegraphics[width=\columnwidth]{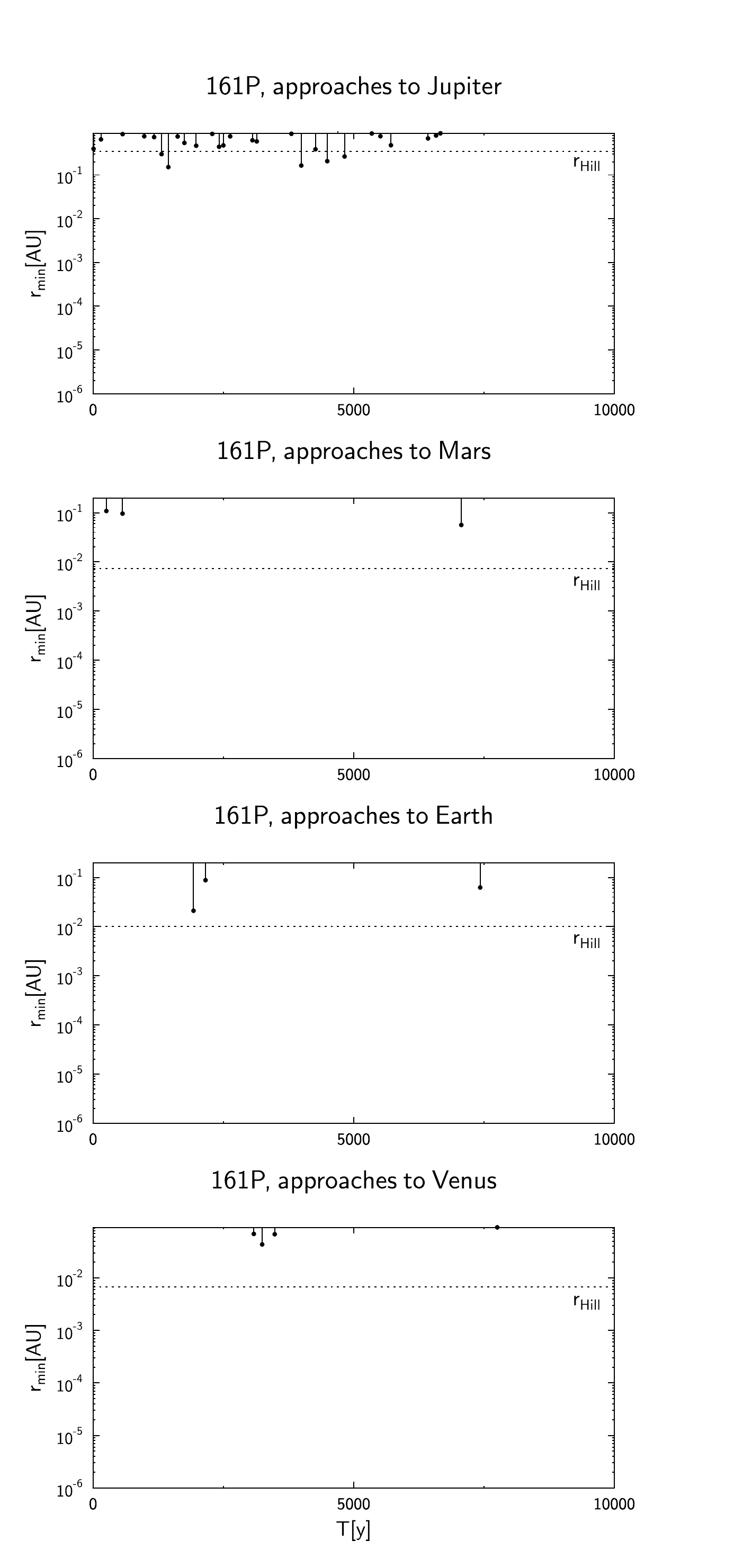}
\caption{Close approaches of of 161P due to Jupiter, Mars, Earth, and Venus. \label{appr_161P}}
\end{figure}

The comet 161P is an object with a different dynamical behaviour from its predecessor (333P) described in this work. Thanks to its near-polar orbit it is a little more stable. The nominal LT value is about 650 years. It is most often approaching Jupiter, but there are also sporadic cases of approaches to Earth, Mars, and Venus. As shown in Fig. \ref{evolution_161}, a possible scenario is that the perihelion distance will reach values close to 1 au and the comet will be in a very eccentric orbit. In this case, as well, cometary effects cause significant changes in orbital elements (Fig. \ref{ngr161}). The close approaches to planets are shown in Fig. \ref{appr_161P}.

To maintain the consistency of the results, we determined the spectrum of LT values for 161P, using the same integration parameters as for 333P. We, therefore, examined 100 clones near the nominal solution, taking into account the maximum possible errors of non-gravitational parameters $A_1$ and $A_2$.

\begin{figure}
\includegraphics[width=\columnwidth]{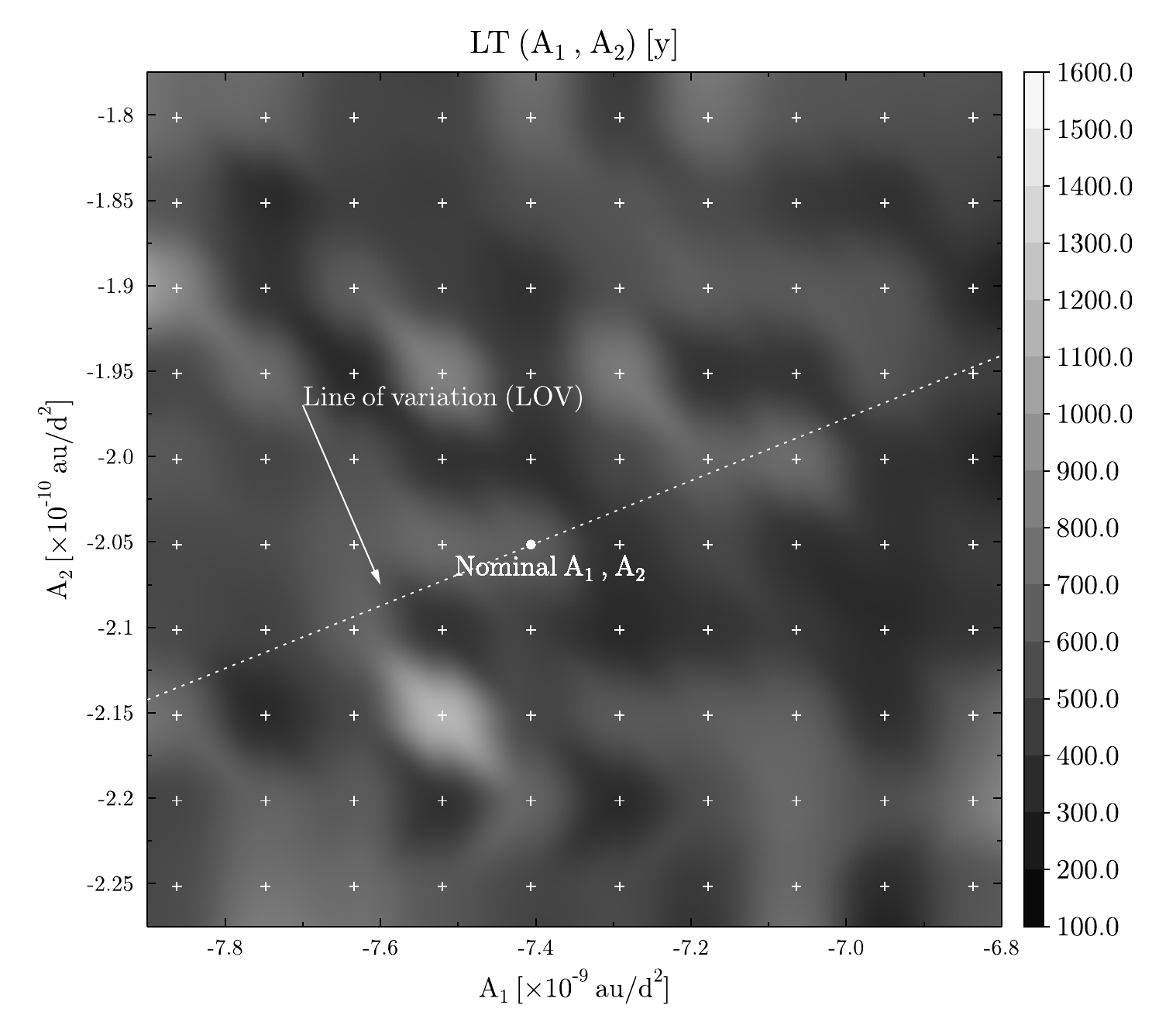}
\caption{Dependence of LT on non-gravitational acceleration parameters (model: $A_1$, $A_2)$ for comet 161P.\label{161lt}}
\end{figure}

\begin{figure}
\includegraphics[width=\columnwidth]{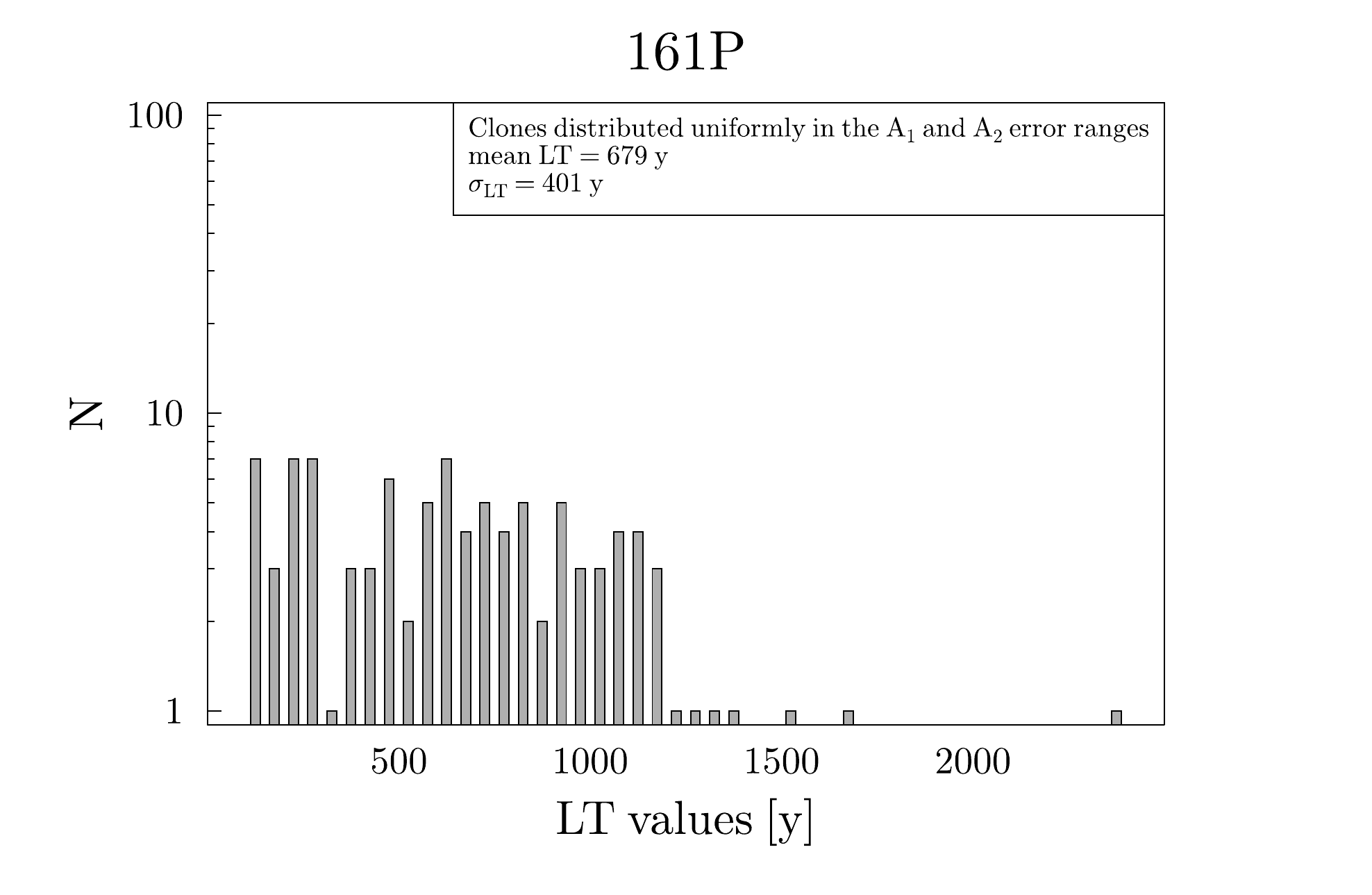}
\caption{Histogram of LT values for comet 161P. Result of the numerical integration of 100 clones, distributed in the error ranges of $A_1$, $A_2$. \label{161lt-hist-matrix}}
\end{figure}

\begin{figure}
\includegraphics[width=\columnwidth]{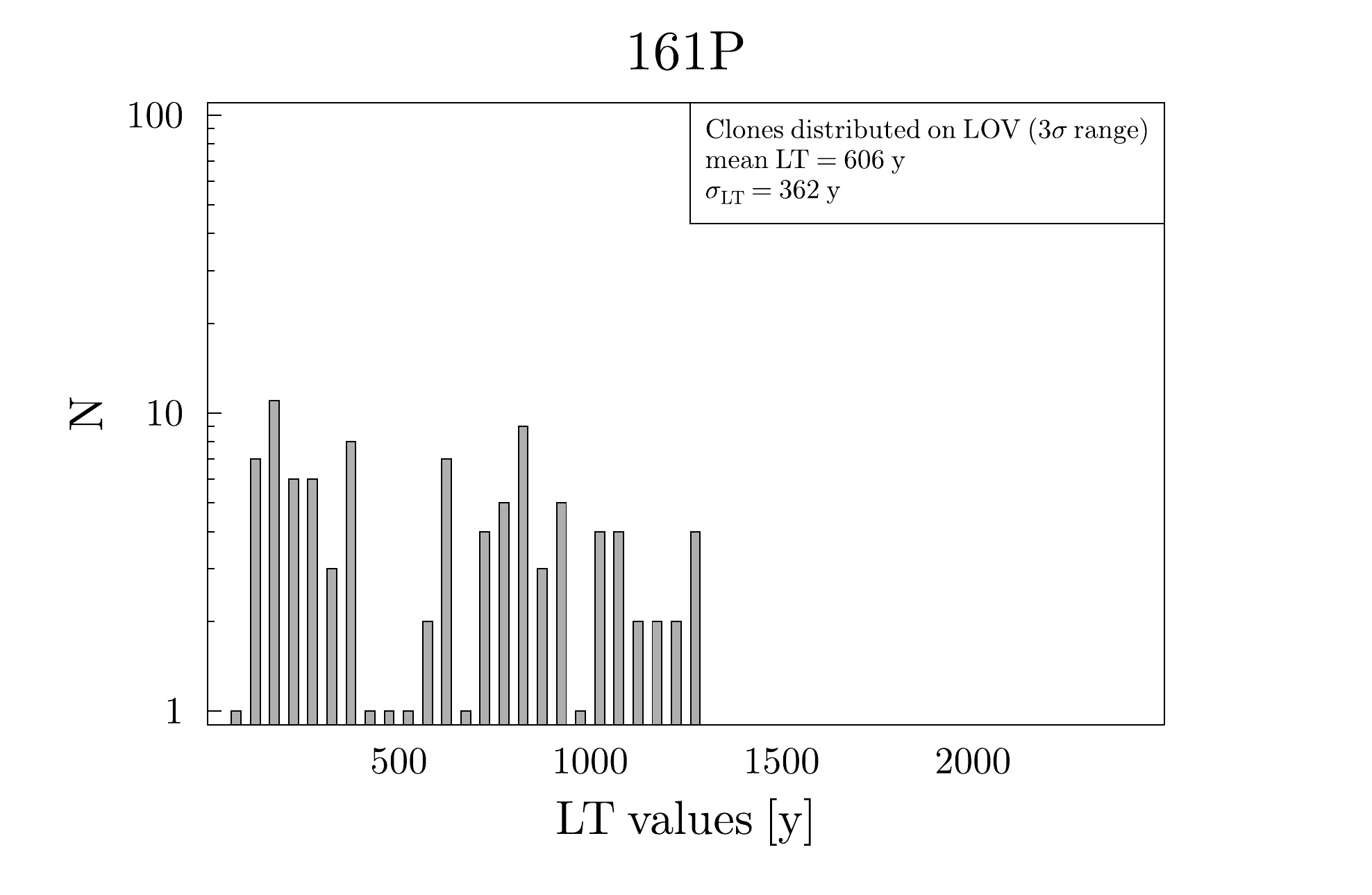}
\caption{Histogram of LT values for comet 161P. Result of the numerical integration of 100 clones, distributed on the LOV, close to the nominal solution. \label{161lt-hist-lov}}
\end{figure}

Similarly, as in the previous case, LT values depending on parameters $A_1$ and $A_2$ for the comet 161P are shown in Fig. \ref{161lt}. In cloning, we applied two strategies again: in the first, the whole range of $A_1$ and $A_2$ errors for 100 clones was investigated (Fig. \ref{161lt-hist-matrix}) and in the second, 100 clones were placed along the LOV (Fig. \ref{161lt-hist-lov}). For comet 161P, the resulting LT distributions are slightly different, but statistical parameters such as mean value and dispersion are similar. Even in this case, non-gravitational effects cause significant changes in LT values, which can range from several hundred up to above a thousand years.

\begin{table*}
\footnotesize\addtolength{\tabcolsep}{-4pt}
\caption{Retrograde comet 161P/Hartley-IRAS: orbital elements (JD 2458600.5) and selected physical data\label{table-161p}.}
\begin{center}
\begin{tabular}{|c|c|c|c|c|c|c|c|}
\hline
Obj.   name     &        $a[au]$    &       $e$     &  $i_{2000}[deg]$ &     $\Omega_{2000}[deg]$    &       $\omega_{2000}[deg]$    &  $M[deg]$& no. of  \\
 & & & & & & & obs. used \\
\hline
161P Hartley/IRAS &  7.727 & 0.835 & 95.676 & 1.362 &  46.908 & 232.799 & 361 \\
1-$\sigma$ rms &  5.539E-06 & 6.710E-07 &  3.358E-05 &  3.854E-05 &  4.603E-05 &  2.645E-06 &   \\ 
\hline

\multicolumn{8}{|c|}{Physical data:}\\
\hline
\multicolumn{8}{|l|}{JPL NASA Classification: Halley-type comet* [NEO]}\\
\multicolumn{8}{|l|}{Cometary parameters, total magnitude, and its slope: M1=11.4, M2=15.0}\\
\multicolumn{8}{|l|}{Cometary parameters NGR [$AU/d^{2}$]:}\\ 
\multicolumn{8}{|l|}{$A_1$=-7.40623E-9 $\pm$ 4.55915E-10  (radial)}\\
\multicolumn{8}{|l|}{$A_2$=-2.05151E-10 $\pm$ 2.00006E-11 (transverse)}\\
\multicolumn{8}{|l|}{Earth MOID ({\it minimum orbit intersection distance}):  0.447763 $AU$}\\
\hline
\end{tabular}
\end{center}
\end{table*}

\section{Conclusions}

Our results appear to be sensitive to the Yarkovsky effect. This force can potentially play an important role,
especially for timescales of $10^6$ y, since its presence could even slightly lengthen the dynamical lifetimes. This conclusion applies not only to the previously received lifetimes but also to the LT values. This confirms our previous results \citep{Kankiewicz2017,Kankiewicz2018}, as well as our assumption that both thermal and cometary effects can stabilize the retrograde orbits.

The decision to use the non-gravitational effect model for an asteroid (Yarkovsky effect) or comet (cometary activity) is crucial for subsequent conclusions. For example, the drift in the semimajor axis for a comet approaching the Earth, such as the 333P, can be at least an order of magnitude greater after taking into account a cometary activity rather than the Yarkovsky effect. Similar numerical calculations carried out with the comet 161P also show that the cometary activity can be significant in the problem studied in this paper. Therefore, these effects influence the behaviour of LCI and LTs.

Comparing the results with other bodies of the Solar System, it can be said that most of the LTs obtained is in the range of $10^4$ to $10^5$ years and these are low values compared to the main belt asteroids. In the AstDys database\footnote{\href{https://newton.spacedys.com/~astdys2}{https://newton.spacedys.com/\~{}astdys2}} \citep{Knezevic2003}, most of numbered non-resonant asteroids have LTs of $10^6$ or more. On the other hand, NEAs have even shorter LTs in the range of hundreds or even decades. Although we initially assumed that the LT values received would be more varied, the method used allowed us to distinguish orbits that are more chaotic than others quantitatively. In comparison to the concept of median dynamical lifetime, this allows us to estimate chaos in a way that is independent of what was previously used. All of this allows us to assess whether the relatively short lifetime in a retrograde orbit (compared to the behaviour of similar objects) is induced by chaos. The question of whether the chaos indicators are strongly correlated with the dynamical lifetimes of asteroids has been brought up several times in the past \citep{Soper1990,Murison1994,Morbidelli1996}. In the absence of a clear correlation between the estimated LT and median lifetime ($\tau$) values, we may be dealing with the so-called phenomenon of stable chaos. Even though the region is chaotic and objects barely differ very much in terms of chaoticity in the sense of Lyapunov, we have more or less stable orbits among our sample. An essential role in this issue may be played by specific RMMRs that change the stability of particular retrograde orbits. The influence of the same resonances that we found for individual objects on our list has also been recently confirmed by \cite{Li2019}. Some of these have been identified by the above authors recently, so their investigation will probably be a new challenge. 

At the current level of model sophistication, precision, and availability of observational data, an update to the database of physical properties of the retrograde asteroids is necessary. This would certainly help to clarify the dynamical models and alleviate ambiguities that otherwise hinder us from drawing new conclusions regarding the scenarios of origin and the nature of these bodies. Luckily, there are opportunities to use future observations, both from space missions as well as ground-based campaigns conducted on professional and amateur telescopes. These observations may, for example, include hunting for new comets, combining dense and sparse photometric results, or determining of spin axis coordinates and rotational properties. Continuing our project, along with the improvement of the model and new data, we hope to explore more detailed dynamical scenarios for individual bodies in retrograde orbits. Future progress in this work will, in particular, make it possible to explain how these orbital inclinations may have evolved to the current extreme values.

\section*{Acknowledgements}
The authors would like to thank the reviewer for his constructive comments, which improved the quality of this work. 

Simulations in Section \ref{model_section} made use of the REBOUND code which is freely available at \verb+http://github.com/hannorein/rebound+.

\newpage
\bibliography{biblio}
\bibliographystyle{aa}

\appendix

\end{document}